\documentclass[10pt,journal,compsoc]{IEEEtran}

\ifCLASSOPTIONcompsoc
  \usepackage[nocompress]{cite}
\else
  \usepackage{cite}
\fi

\usepackage{amsmath,amsthm,graphicx,cite,booktabs,amssymb,gensymb, color, xcolor, fancyhdr,float,perpage,tikz, colortbl}
\usepackage[linesnumbered,vlined,ruled]{algorithm2e}
\usepackage{algorithmic,array,enumerate,rotating}
\usepackage{subfigure,epsfig,dblfloatfix}
\usepackage{multirow}
\usepackage{hyperref}
\usetikzlibrary{matrix}
\hypersetup{
	colorlinks   = true,
	linkcolor    = blue,
	citecolor    = blue
}

\makeatletter

\renewcommand{\@algocf@capt@plain}{above}

\newcommand{\norm}[1]{\left\lVert #1 \right\rVert}

\usepackage{pifont}% http://ctan.org/pkg/pifont
\newcommand{\cmark}{\ding{51}}%
\newcommand{\xmark}{\ding{55}}%

\definecolor{LightGray}{gray}{0.85}
\definecolor{Gray}{gray}{0.65}

\newcommand\bdr[1]{{\textbf{#1}}}
\newcommand\bdb[1]{{\underline{#1}}}

\newcommand{\sd}{\textcolor{black}}
\newcommand{\dk}{\textcolor{black}}

\makeatother
\def\bsx{{\boldsymbol{x}}}
\def\bsy{{\boldsymbol{y}}}
\def\bsX{{\boldsymbol{X}}}
\def\bsY{{\boldsymbol{Y}}}
\def\bse{{\boldsymbol{e}}}
\def\bsH{{\boldsymbol{H}}}
\def\bsI{{\boldsymbol{I}}}
\def\bsJ{{\boldsymbol{J}}}
\def\bsK{{\boldsymbol{K}}}
\def\bsL{{\boldsymbol{L}}}
\def\bsP{{\boldsymbol{P}}}

\def\bsB{{\boldsymbol{B}}}
\def\bsC{{\boldsymbol{C}}}
\def\bsE{{\boldsymbol{E}}}
\def\bsG{{\boldsymbol{G}}}
\def\bsO{{\boldsymbol{O}}}
\def\bsP{{\boldsymbol{P}}}
\def\bsR{{\boldsymbol{R}}}
\def\bpsi{{\boldsymbol{\psi}}}
\def\balpha{{\boldsymbol{\alpha}}}

\begin{document}

% Title.
% ------
\title{DIVA: Deep Unfolded Network from Quantum Interactive Patches for Image Restoration}
% of Image Denoising by Quantum Interactive Patches

\author{Sayantan~Dutta,~\IEEEmembership{Student Member,~IEEE,}
Adrian~Basarab,~\IEEEmembership{Senior~Member,~IEEE,}
Bertrand~Georgeot,
and~Denis~Kouam\'e,~\IEEEmembership{Senior~Member,~IEEE}
% <-this % stops a space
\IEEEcompsocitemizethanks{\IEEEcompsocthanksitem S. Dutta is with the IRIT, Universit\'e de Toulouse, CNRS, Toulouse INP, UT3, Toulouse, France and the Laboratoire de Physique Th\'eorique, Universit\'e de Toulouse, CNRS, UPS, France. E-mail: sayantan.dutta@irit.fr.%\protect\\
\IEEEcompsocthanksitem A. Basarab is with the Universit\'e de Lyon, INSA‐Lyon, Université Claude Bernard Lyon 1, UJM-Saint Etienne, CNRS, Inserm, CREATIS UMR 5220, U1294, F‐69621, Villeurbanne, France.
\IEEEcompsocthanksitem B. Georgeot is with the Laboratoire de Physique Th\'eorique, Universit\'e de Toulouse, CNRS, UPS, France.
\IEEEcompsocthanksitem D. Kouam\'e is with the IRIT, Universit\'e de Toulouse, CNRS, Toulouse INP, UT3, Toulouse, France.}% <-this % stops a space
%\thanks{Manuscript submitted August 1, 2022.}
\thanks{(Corresponding author: Sayantan~Dutta).}
}

\markboth{Submitted}%
{Dutta \MakeLowercase{\textit{et al.}}:DIVA: Deep Unfolded Network from Quantum Interactive Patches for Image Restoration}

\IEEEtitleabstractindextext{%
\begin{abstract}
This paper presents a deep neural network called DIVA unfolding a baseline adaptive denoising algorithm (De-QuIP), relying on the theory of quantum many-body physics. Furthermore, it is shown that with very slight modifications, this network can be enhanced to solve more challenging image restoration tasks such as image deblurring, super-resolution and inpainting. Despite a compact and interpretable (from a physical perspective) architecture, the proposed deep learning network outperforms several recent algorithms from the literature, designed specifically for each task. The key ingredients of the proposed method are on one hand, its ability to handle non-local image structures through the patch-interaction term and the quantum-based Hamiltonian operator, and, on the other hand, its flexibility to adapt the hyperparameters patch-wisely, due to the training process.
\end{abstract}

}

\IEEEdisplaynontitleabstractindextext

\IEEEpeerreviewmaketitle

\ifCLASSOPTIONcompsoc

%\pagenumbering{gobble}

%
\maketitle
%

%\begin{abstract}

%This paper presents a deep \dk{neural} network unfolding \dk{ a baseline  adaptive denoising algorithm (De-QuIP), relying on the theory of quantum many-body physics}. Furthermore, it is shown that with very slight modifications, this network can be enhanced to solve more challenging image restoration tasks such as image deblurring, super-resolution and inpainting. Despite a compact and interpretable (from a physical perspective) architecture, the proposed deep learning network outperforms several recent algorithms from the literature, designed specifically for each task. The key ingredients of the proposed method are \dk{on the one hand, }its ability to handle non-local image structures through the patch-interaction term and the \dk{quantum-based} Hamiltonian operator, and \dk{on the other hand,} its flexibility to adapt the hyperparameters patch-wise, due to the training process.
%%\vspace*{-5pt}
%\end{abstract}

%\begin{IEEEkeywords}
%Quantum many-body interaction, \dk{Schr\"odinger equation} unfolding, deep learning, image restoration, quantum image processing.
%%\vspace*{-16pt}
%\end{IEEEkeywords}

%%\vspace*{-18mm}
\section{Introduction}
\label{sec:intro}

\IEEEPARstart{R}{estoring} a high-quality image from a degraded observation is a classic \sd{but still} major challenge in imaging applications, such as medical imaging, remote sensing, low-level vision, surveillance, to cite few. Such a degradation process can be formulated as
%\begin{equation}
$\bsY = \bsO \bsX + \bse$,
%\end{equation}
where, $\bsY$ and $\bsX$ denote the low quality observation and the image of interest, respectively, $\bsO$ denotes the degradation operator, and $\bse$ is associated with an \dk{additive  noise}. The goal is to recover the underlying high-quality \dk{image} $\bsX$ from the \dk{observation} $\bsY$. Depending on the degradation operator $\bsO$, different restoration problems occur. For example, if $\bsO$ is the identity operator, the resulting problem is an image denoising \cite{donoho1994ideal, Elad2006image, Dabov2007Image, Dong2013nonlocal} problem. If $\bsO$ is a blurring operator then restoration becomes a deblurring \cite{Danielyan2012bm3d, afonso2010fast, chan2013constrained, almeida2013deconvolving, Chen2016Compressive},
%yang2010fast,
or a super-resolution (SR) task \cite{Gao2012image, Mudunuri2016low, Zhao2016fast} if $\bsO$ includes a subsampling operator. In practice, estimating $\bsX$ from $\bsY$ by mitigating the effect of the degradation operator $\bsO$ is a challenging ill-posed inverse problem. Over the past few decades, image restoration techniques have been extensively studied, yet remain an active field of research.

Traditionally, the restoration process is framed as a model-based optimization problem from a Bayesian perspective, in which the desired solution is obtained \dk{by} minimizing the sum \dk{of} a
%prior/
\sd{regularization and a data fidelity term.} Over the time, numerous model-based regularizers have been introduced in the literature including total variation \cite{Osher2005Iterative}, sparsity-based transformations \cite{donoho1994ideal}, sparse models \cite{Elad2006image, Aharon2006an, Dong2013nonlocally}
%Mairal2008sparse,
and spatial filtering \cite{tomasi1998bilateral, Durand2002fast}, in particular non-local self-similarity (NLSS) filters \cite{buades2005review, Mairal2009nonlocal}, anisotropic diffusion filters \cite{Perona1990scale, Yongjian2002speckle}, guided filters \cite{He2013guided}, etc.
%buades2008nonlocal, tasdizen2009principal, Kou2015gradient
In particular, non-local regularization approaches \cite{Dong2011image, Sun2011Gradient, Dong2013nonlocally, Teodoro2016image} blending the NLSS and low-rank regularity, such as BM3D \cite{Dabov2007Image},
% , dabov2009bm3d
NLM \cite{buades2005review}, LSSC \cite{Mairal2009nonlocal}, NCSR \cite{Dong2013nonlocally},
% NLM \cite{tasdizen2009principal}, WNNM \cite{gu2017weighted}, 
etc., have been extensively discussed due to their state-of-the-art restoration \dk{performances}. Integration of non-local information in the process of retrieving a particular region of the image is the key to the success of the NLSS models.
%\dk{ REMOVE THIS SENTENCE : More recently, the promising denoising performances of these NLSS models have conspired to plug them into image recovery problems as an off-the-shelf denoising schemes, where popular algorithms are Plug-and-Play \cite{venkatakrishnan2013plug} and regularization by denoising \cite{romano2017little}.}
In general, the model-based approaches are fairly successful in tackling a variety of image retrieval tasks, including proper interpretation of their roles. However, these schemes require conducting a costlier computation process and manual tuning of several hyperparameters, which are the primary challenges of these strategies.

\dk{Based on deep convolutional neural networks (CNN), deep learning (DL)-based strategies brought an alternative to the well-established model-based methods to counter such imaging problems.  DL algorithms \cite{vincent2010stacked, Burger2012image, he2016deep, Zhang2018FFDNet, Nan2020variational, Zhang2017learning, Zhang2021residual, Dong2019denoising, Zha2021Triply, Zha2022Low}}
% Kiranyaz2020operational, Malik2021self, Dong2014learning,
achieved state-of-the-art performances in recent years by learning the mapping functions from observed degraded or low-resolution (LR) images to the original or high-resolution (HR) images. CSF \cite{schmidt2014shrinkage}, TNRD \cite{Chen2017trainable}, DnCNN \cite{Zhang2017beyond}, Super-ONN \cite{malik2021image},
%RGDN \cite{Gong2020learning}, RCAN \cite{Zhang2018image},
DWDN \cite{Dong2020deep}, SRCNN \cite{Dong2016image}, RDN \cite{Zhang2021residual}, DRLN \cite{Anwar2022densely}, etc., are some of the well-known DL networks with proven efficiency in image restoration over the conventional model-based approaches, exploiting a training dataset in the learning process. However, training a CNN is not straightforward. The performance largely depends on the number of layers, the kernel size and the learning rate. Deeper network structures may provide better results but exponentially increase the training complexity \cite{he2016deep}. Thus, network structures are in most cases determined empirically, which makes them suffer from a lack of interpretation of their true functionality.

Benefiting from CNN's powerful representation ability, a new concept, known as unfolding \cite{Gregor2010learning}, gathering the advantages of both model and DL-based approaches, is currently \sd{gaining more attention
% in imaging domain 
due to its explanatory properties.} The main idea of such frameworks is to construct a DL network starting from a classical algorithm. This approach has recently been successfully explored in the literature, leading to superior restoration performance over the classical peer, such as BM3D-NET \cite{Yang2018bm3dnet}, LKSVD \cite{Scetbon2021deep}, FBPConvNet \cite{Jin2017deep}, DRED-DUN \cite{Kong2022deep}, CORONA \cite{Solomon2020deep}, USRNet \cite{zhang2020deep}, to cite few.

% NLMNet \cite{lefkimmiatis2017non}, deep-ULM \cite{van2020deep}, USPLS \cite{koo2022bayesian}, DUBLID \cite{li2019deep}, UWMMSE \cite{Chowdhury2021unfolding}, L-RFPI \cite{Khobahi2020model},

% With the objective of gathering the advantages of both model and DL-based approaches, a new concept, known as unfolding \cite{Gregor2010learning}, is currently gaining more attention due to its explanatory properties while benefiting from CNN's performance. Exploring this aspect, few works have recently been proposed in the literature significantly outperforming their classical peer,

\dk{In this work}, we advocate novel CNN architectures for image restoration problems, unfolding our recently introduced quantum mechanics-based adaptive denoising algorithm called
%Denoising by Quantum Interactive Patches
\sd{De-QuIP} \cite{dutta2021image, dutta2022novel}. In the last decade, quantum mechanics-based frameworks have been explored in the field of image processing and analysis, such as the single-particle quantum theory, especially for image segmentation \cite{aytekin2013quantum, youssry2015quantum}, denoising \cite{kaisserli2015novel, dutta2021quantum}, deblurring \cite{dutta2021plug, dutta2021poisson} or others \cite{Altmann2018quantum}. Despite the promising performances, these single-particle-based frameworks cannot benefit from the structural features of the image like NLSS algorithms. In contrast, De-QuIP is based on the theory of many-body quantum systems, where each image patch behaves like a single particle system and interacts with its neighbors. This phenomenon of interaction preserves the image similarity/features from a local neighborhood. Indeed, absorption of this concept of interaction in De-QuIP brings an intrinsic non-local structure to the algorithm that notably enhances the denoising performance and has been extensively \sd{studied in \cite{dutta2022novel}.}
% studied in our previous works \cite{dutta2021image, dutta2022novel}.
Despite its interesting performances, De-QuIP struggles with costly computational processes (\textit{e.g.}, hyperparameters tuning and eigenvalue decomposition) like many other model-based schemes, which may limit its practical use.

In this paper, we introduce a novel DL network unfolding the baseline De-QuIP algorithm, denoted as \dk{DIVA (Deep denoising by quantum InteractiVe pAtches)} for image denoising problem. We further extend the network architecture to conduct a general image restoration task. The inclusion of the quantum interaction theory brings a non-local structure to the proposed CNN architecture. Indeed, in our depicted DL models, the fundamental aspects of quantum theory from the baseline De-QuIP algorithm are essentially preserved. Furthermore, the DL model efficiently resolves the hyperparameter tuning problem of the original De-QuIP scheme, harnessing the power of back-propagation. The integration of the key attributes of DL and quantum theory significantly enhances the functionality of our proposed networks due to their intrinsic versatility and enables our models to exhibit state-of-the-art performances for several restoration tasks.

An initial illustration of this work was presented as a conference report that portrays preliminary results on Gaussian denoising \cite{dutta2022deep}. Herein, we extend our preliminary model to a robust generalized formalism by incorporating additional contents in significant ways:
(i) we extend the initially proposed DL model, primarily designed for denoising, to more complex image restoration tasks such as deblurring, super-resolution and inpainting, with a resilient generalized network architecture;
(ii) we conduct a detailed investigation regarding the network diagram and add considerable analysis of the incorporated quantum \sd{background}, tunable parameter number, and run time;
(iii) we report a comprehensive survey of image restoration performance against benchmark methods for various imaging problems.

The remainder of the paper is organized as follows. Sec.~\ref{sec:related_work} reminds briefly the concepts of the baseline De-QuIP algorithm \dk{for self-consistency reasons}. Sec.~\ref{sec:deep_aechi} first presents the proposed \dk{DIVA} network for denoising and then extends it to an advanced model for other imaging tasks. The experimental settings and extensive evaluations are reported in Sec.~\ref{sec:expe_results}.
\sd{Sec.~\ref{sec:discuss} outlines the overall remarks and possible future perspectives.}
%We discuss the overall remarks and possible future perspectives in Sec.~\ref{sec:discuss}.
Finally, Sec.~\ref{sec:conclusion} draws the conclusions.

\section{Brief Review of Quantum-Interactive-Patches-Based Denoising}
\label{sec:related_work}

To facilitate the understanding of the proposed method, we briefly revisit the baseline De-QuIP algorithm for image denoising \sd{and its main properties}.
%with its pros and cons.

%\vspace{-4mm}
\subsection{The De-QuIP Scheme}
\label{sec:dequip}
%%\vspace*{-2pt}
Built on an underlying nonlocal architecture, De-QuIP \cite{dutta2021image, dutta2022novel} offers an adaptive way of image denoising based on the theory of quantum many-body interaction. The theory of quantum many-body physics \sd{subscribes}
%narrates the attributes of 
many-body quantum systems, where inevitably particle-to-particle interactions emerge. De-QuIP provides a framework for extending this concept of interaction to imaging problems. Effectively, De-QuIP divides an image into small patches, and each image patch \sd{acts} as a single-particle system while interacting with its neighbors, i.e., with neighboring patches, inside the whole image, similarly to a many-body system. Indeed, these interactions between neighbors reflect their mutual similarities that enhance the denoising performance of De-QuIP significantly.

Similar to any denoising method, the goal is to estimate the underlying clean image $\bsX \in \mathbb{R}^{M \times N}$ from a noisy observation $\bsY \in \mathbb{R}^{M \times N}$. The respective vectorized representations of $\bsX$ and $\bsY$ are denoted by $\bsx \in \mathbb{R}^{MN}$ and $\bsy \in \mathbb{R}^{MN}$ in lexicographical order. \sd{Based on the many-body quantum physics, the primary idea of De-QuIP algorithm is to construct an adaptive transformation using the wave solutions of the Schr\"odinger equation $H \bpsi (z) = E \bpsi (z) $, where the wave function $\bpsi(z)$ describes a particle with energy $E$ in a potential $V$, $z$ being the spatial coordinate. In a many-body system, denoting by $I$ the interaction, the Hamiltonian operator is $H = -(\hbar ^2/2m)\nabla^2 + V + I$, where $\nabla$ and $(\hbar ^2/2m)$ are respectively the gradient operator and a function of the Planck's constant (this function acts as a hyperparameter in this formalism). For this patch-based imaging scheme, the potential $V$ is represented by the original pixels' values of the image patch and the patch-similarity measures act as the interaction $I$. The set of eigenvectors of the Hamiltonian operator gives the adaptive transformations for the respective patch.
Thus, for a system with multiple particles, the Hamiltonian operator $\bsH_a$ for the $a$-th patch is defined by:
%%\vspace{-2mm}
\begin{equation}
%%\vspace{-2mm}
\bsH_a = -(\hbar ^2/2m)\nabla^2 + \bsJ_a + \bsI_a,
\label{eq:hamiltonian}
\end{equation}
where $\bsJ_a$ and $\bsI_a$ are respectively the pixels' values and interaction term for the $a$-th patch. The corresponding set of eigenvectors $\bsB_a$ of $\bsH_a$ acts as the quantum adaptive basis for the $a$-th patch.
% The De-QuIP architecture consists of the following the key steps:
The key steps of De-QuIP algorithm are as follows.}

\textit{Patch extraction:}
The patch extraction step primarily uncoils small patches from the observed image and assimilates their neighbors into their respective local groups. Let us denote by $\bsJ_a$ a patch of size $n^2$ whose upper-left pixel position is $a$, and by $\Omega$ the set containing all such patches extracted from the image $\bsy$. For all $\bsJ_a \in \Omega$, one creates a window of size $W \times W$ centered on $\bsJ_a$ and accumulates all patches inside the window in a set denoted by $S_{\bsJ_a}$ to create local groups.

%Let us denote by $\Omega$ the set containing all the patches of size $n^2$ extracted from the image $\bsy$, and by $\bsJ_a$ a patch whose upper-left pixel position is $a$.

% brings a nonlocal 

\textit{Total interaction:}
The goal of the interaction step is to preserve local structures/similarities by exploiting the local groups through a notion akin to the interactions in quantum mechanics. This step computes the interactions $\bsL_{ab}$, for all $\bsJ_b \in S_{\bsJ_a}$ and all $\bsJ_a \in \Omega$, using power laws of physics \cite{dutta2022novel}, \textit{i.e.}, interaction is linearly proportional to the pixel-wise difference $\bsK_{ab}^k = | \bsJ_a^k - \bsJ_b^k |$ for $k = 1,\cdots, n^2$ and inversely proportional to the square of the Euclidean distance $D_{ab}$ between the patches. Summing over $b$ gives the total interaction for the $a$-th patch
%%\vspace{-3.5mm}
\begin{equation}
%%\vspace{-2.8mm}
\bsI_a = p \sum_b \bsL_{ab} = p \sum_b \bsK_{ab} / D_{ab}^2, ~~\forall \bsJ_a \in \Omega.
\label{eq:powerlaw}
\end{equation}
In this construction the proportionality constant $p$ acts as a hyperparameter.

\textit{Hamiltonian operator and adaptive basis:}
This step formulates the energy or Hamiltonian operators of the extracted patches by incorporating their total interaction with their neighbors in the local group using \eqref{eq:hamiltonian}.
The associated set of eigenvectors $\bsB_a$ of the Hamiltonian operator $\bsH_a$ operates as the adaptive basis for the current image patch $\bsJ_a$.

%Therefore, mathematically, Hamiltonian operator is defined as follows:
%\begin{equation}
%\bsH_a = -(\hbar ^2/2m)\nabla^2 + \bsJ_a + \bsI_a, ~~\forall \bsJ_a \in \Omega,
%\end{equation}
%where $\nabla$ and $(\hbar ^2/2m)$ are respectively the gradient operator and Planck's constant (the Planck's constant acts as a hyperparameter in this formalism). The corresponding set of eigenvectors $\bsB_a$ of the Hamiltonian operator $\bsH_a$ acts as the quantum adaptive basis for the current image patch $\bsJ_a$.

%%\vspace{-4mm}
\textit{Thresholding:}
The thresholding is processed on the coefficients resulting from patch projections onto their respective adaptive basis. Hence, the noise is attenuated by projecting $\bsJ_a$ onto $\bsB_a$ and performing hard/soft-thresholding $\mathcal{T}$ in energy. Finally, reverse projecting the truncated coefficients reinstates the denoised patch $\hat{\bsJ_a}, \forall \bsJ_a \in \Omega$.

\textit{Patch accumulation:}
This step accumulates all the denoised patches to their original positions and normalizes them to reconstruct the estimated denoised image $\hat{\bsx}$. In the following, the patch extractor operator is denoted by $\bsE$, while the operation of accumulating the patches to form the denoised image is denoted by $\bsE^{-1}$.

% In the De-QuIP framework, the patch interaction phenomenon efficiently preserves the local structures of real images in a local image neighborhood. This preserved spatial information

\sd{In the De-QuIP framework, the preserved spatial information by the patch interaction phenomenon coherently passes through the Hamiltonian operator} to the quantum adaptive basis and enables the algorithm to handle denoising tasks regardless of the noise intensity, statistics and image nature. Its application field is not limited to denoising tasks \cite{dutta2021image, dutta2022novel}, and its efficiency has been illustrated in various imaging problems such as despeckling \cite{dutta2021despeckling} and super-resolution \cite{dutta2022quantum}. Fig.~\ref{fig:Arch_com}(a) depicts the De-QuIP architecture, where interaction, proportionality constant, adaptive basis and thresholded coefficients are denoted by $\bsL$, $\bsP$, $\bsB$ and $\bsR$ respectively.

%\vspace{-4mm}
\subsection{Shortcomings of De-QuIP}
\label{sec:short_dequip}
%%\vspace{-1.5mm}
The major challenge of De-QuIP is its high computational cost of tuning the hyperparameters $p$, $(\hbar ^2/2m)$ and energy threshold. In \cite{dutta2022novel}, the influence of these hyperparameters and strategies to optimize their values were discussed. Despite some rules that guided the choice of the hyperparameters, they remain general for the whole image and are not optimized to be applied patch-wise.
%Thus, all the extracted patches from one image are assigned the same hyperparameters.
\sd{Although De-QuIP demonstrates favorable outcomes despite these drawbacks, the intrinsic non-local architecture of the algorithm raises an obvious question of assigning patch-dependent hyperparameter values, which can further enhance the adaptability of the model. However, manually tuning all the hyperparameters separately for each patch is practically impossible. The power of DL architecture removes this barrier by involving many parameters that can be learned during the training process.}
%using different hyperparameter values for different patches. The adaptability of De-QuIP can be further enhanced by these patch-dependent hyperparameters. However, manually tuning all the hyperparameters separately for each patch is practically impossible. The power of deep learning architecture removes this barrier by involving many parameters that can be learned during the training process.

Another challenge of De-QuIP is the computationally-expensive task of adaptive basis vector computation from the Hamiltonian operator. Furthermore, this adaptive basis is exploited to calculate the projection coefficients, bringing additional computational burden. A deep learning model can bypass all these bottlenecks by directly estimating the projection coefficients with the help of convolutional kernels. The subsequent section focuses on this deep-learning prospect of the De-QuIP algorithm, the main contribution of this paper.

% The high computational cost of tuning the hyperparameters $p$, $(\hbar ^2/2m)$ and thresholding energy is a major drawback of De-QuIP.

% Thus, the same values are assigned to the hyperparameters for all patches extracted from one image. Although De-QuIP exhibits promising performance despite these limitations, it raises the question of using different hyperparameter values for different patches due to the nonlocal structure of the algorithm.

%Although De-QuIP exhibits promising performance under this constraint, it is quite obvious to use different hyperparameter values for different patches due to the nonlocal structure of the algorithm.

%\vspace{-4mm}
\section{Proposed Deep Architectures for Image Restoration}
\label{sec:deep_aechi}
%%\vspace{-1mm}
This section presents deep unfolding strategies for image restoration problems built on the baseline De-QuIP algorithm. Depending on the image degradation operator $\bsO$, various imaging problems arise. If $\bsO$ is an identity operator, the image restoration problem is equivalent to a denoising task, whereas, depending on $\bsO$, it may turn into deblurring, super-resolution or inpainting, addressed herein. In the following, two deep architectures are introduced: the first addresses denoising, and the second more complex image restoration tasks. The first proposed network, referred to as \dk{DIVA}, is a direct translation of the baseline De-QuIP algorithm into a deep learning model to handle denoising. To handle non-identity degradation operators $\bsO$, \dk{DIVA} architecture is slightly modified and denoted by \dk{DIVA} advanced (\dk{DIVA}-A). The subsequent subsections illustrate these two network architectures.

% add figures  --------------------------------------
\begin{figure}[t!]
\centering
\includegraphics[width=0.43\textwidth]{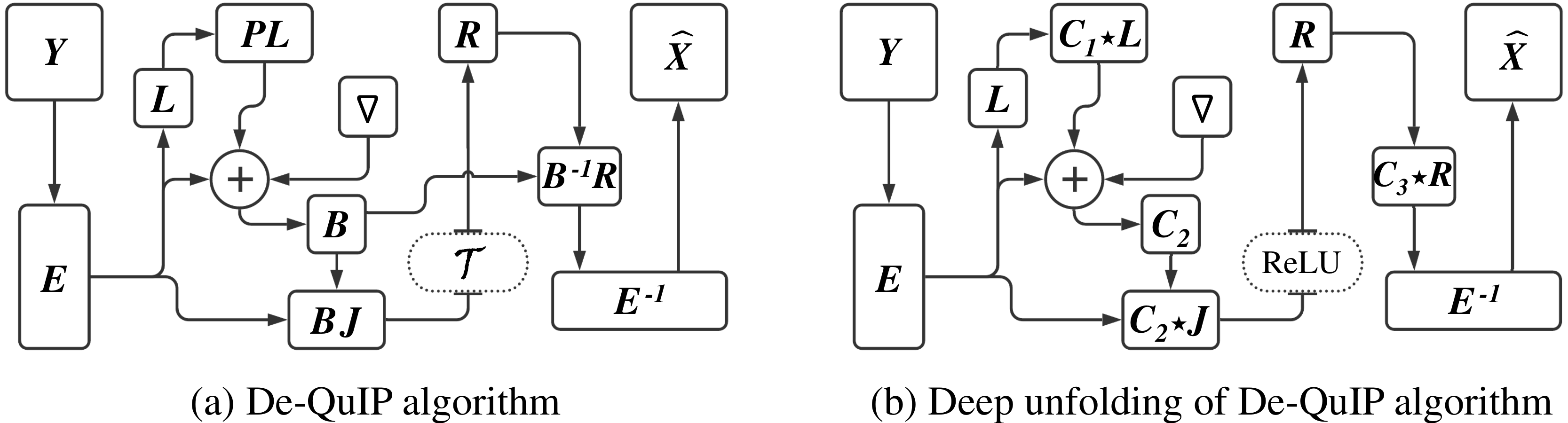}
%%\vspace*{-3.8mm}
\caption{Architectural comparison between De-QuIP and its DL counterpart.}
%\vspace*{-4mm}
\label{fig:Arch_com}
\end{figure}

% add figures  --------------------------------------
\begin{figure*}[h!]
\begin{centering}

\subfigure[Proposed DIVA network for image denoising.]{
\label{subfig:deep_dequip_layers}
\includegraphics[width=1\textwidth]{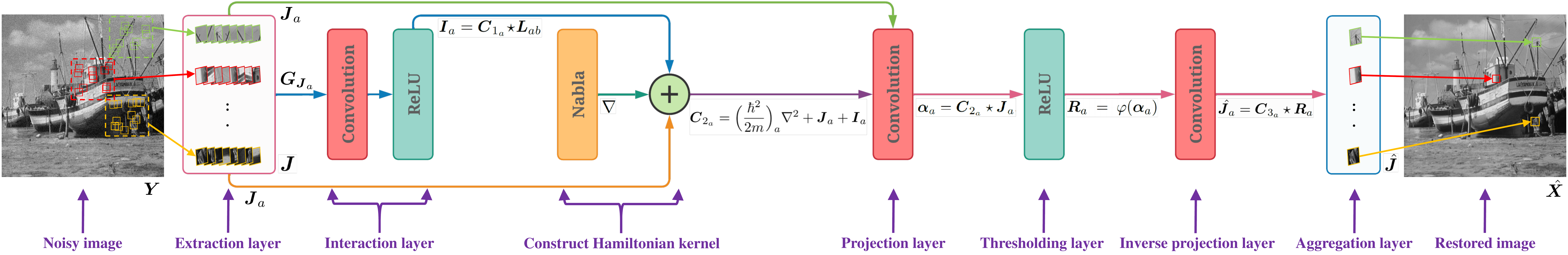}}\\ %\vspace*{-2mm}
\subfigure[Proposed DIVA-A network for image restoration.]{
\label{subfig:deep_dequipA_layers}
\includegraphics[width=1\textwidth]{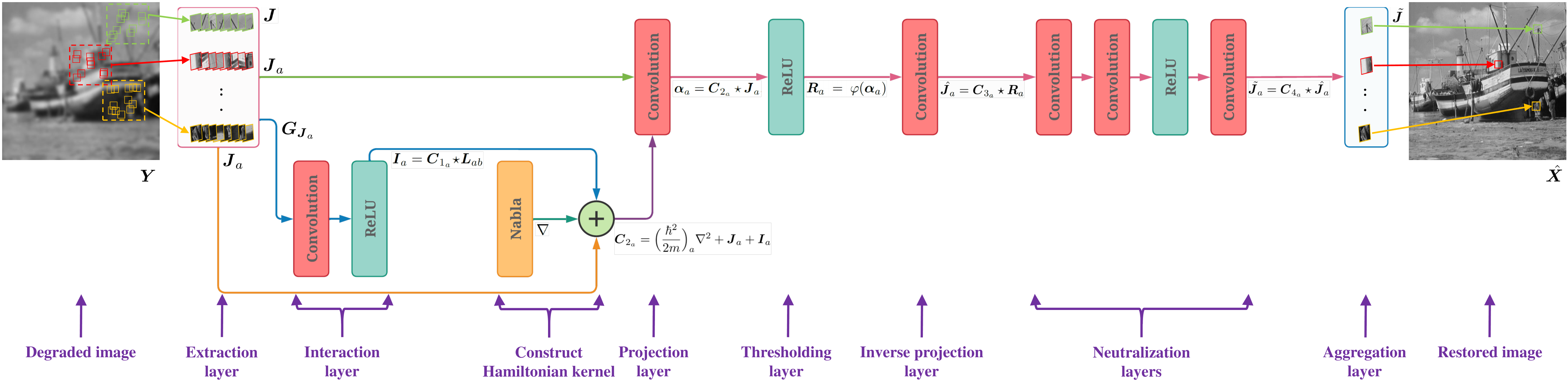}}

\end{centering}
%\vspace*{-2mm}
\caption{The architectures of the proposed deep learning models. The corresponding operations for a patch $\bsJ_a$ are indicated next to each block.}
%\vspace*{-4mm}
\label{fig:deep_model_layers}
\end{figure*}

%\vspace{-4mm}
\subsection{Proposed DIVA Architecture}
\label{sec:deep_dequip}
%%\vspace{-1.3mm}
\dk{The main idea behind the proposed unfolding strategy is to replace the matrix multiplication steps in De-QuIP by convolution layers. The analogy between the original algorithm and its unfolded version is illustrated in Fig.~\ref{fig:Arch_com}.}
The proposed \dk{DIVA} network primarily stands upon eight main pillars.

\textit{Extraction layer:}
Similar to the De-QuIP algorithm, the extraction layer in \dk{DIVA} assembles all patches from a local window of size $W \times W$ centered at $\bsJ_a$ in a local patch-group denoted as $S_{\bsJ_a}, \forall \bsJ_a \in \Omega$. Let the cardinality of $S_{\bsJ_a}$ be $\kappa$, $\forall \bsJ_a \in \Omega$ and $\zeta$ be the cardinality of $\Omega$. The patch extraction operation from the local window can be defined as a matrix multiplication by $\bsE_{\bsJ_a} \in \mathbb{R}^{n^2 \kappa \times MN}$ for each $\bsJ_a$. Therefore, mathematically, $ \bsG_{\bsJ_a} = \bsE_{\bsJ_a} \bsy$, where $\bsG_{\bsJ_a} \in \mathbb{R}^{n^2 \kappa}$ is the concatenated vectorized local patch group for each $\bsJ_a$. Thus, for the whole image, the patch extractor operator $\bsE \in \mathbb{R}^{\zeta n^2 \kappa \times MN}$ is constructed by concatenating $\bsE_{\bsJ_a} \in \mathbb{R}^{n^2 \kappa \times MN}$ $\forall \bsJ_a \in \Omega$. Finally, $\bsJ_a$ and $\bsG_{\bsJ_a}$ $\forall \bsJ_a \in \Omega$ are concatenated and reshaped to construct matrices $\bsJ \in \mathbb{R}^{\zeta \times n^2}$ and $\bsG \in \mathbb{R}^{\zeta \times n^2 \kappa}$, further considered as inputs for the next layer.

\textit{Interaction layer:}
This layer computes the interactions between patches for each local group $\bsG_{\bsJ_a}$ following the power laws discussed in Sec.~\ref{sec:dequip}. But rather than considering a fixed hyperparameter value $p$ as in \eqref{eq:powerlaw}, for each local group $\bsG_{\bsJ_a}$ a different set of $p_{ab}^k$ values is assigned for each pixel $k$ ($k = 1,\cdots, n^2$) and patch $b$ ($b = 1,\cdots, \kappa; \neq a$) respectively. Therefore, the total interaction can be expressed as
%%\vspace{-2.5mm}
\begin{equation}
%%\vspace{-2mm}
\bsI_a^k = \sum_{b=1, b\neq a}^\kappa p_{ab}^k  \dfrac{\bsK_{ab}^k}{D_{ab}^2} = \sum_{b=1, b\neq a}^\kappa p_{ab}^k \bsL_{ab}^k, \mbox{for each }\bsG_{\bsJ_a}.
\label{eq:interaction}
\end{equation}
In matrix notation, $\bsI_a = \bsP_{ab} \bsL_{ab}$, for each $\bsG_{\bsJ_a}$, where $\bsI_a \in \mathbb{R}^{n^2}$, $\bsP_{ab} \in \mathbb{R}^{n^2 \times n^2 (\kappa-1)}$, and $\bsL_{ab} \in \mathbb{R}^{n^2 (\kappa-1)}$ respectively denote the total interaction for patch $\bsJ_a$, proportionality constant in local group $\bsG_{\bsJ_a}$, and interaction between $\bsJ_a$ and $\bsJ_b$ patches. At this point, the main challenge is to tune the values of $\bsP_{ab}$ so that $\bsI_{a}$ can efficiently preserve the local information and incorporate them into the Hamiltonian. One may note that this process is equivalent to a \dk{ convolution between $\bsL_{ab}$ and} a learnable filter $\bsC_{1_a}$ of appropriate size. Hence, the local operation in the layer is,
%%\vspace{-2.5mm}
\begin{equation}
%%\vspace{-2.3mm}
\bsI_a = \bsC_{1_a} \star \bsL_{ab}, ~~\forall \bsG_{\bsJ_a},
\end{equation}
where $\star$ indicates the convolution product. This convolution layer is followed by a Rectified Linear Unit (ReLU) to truncate the insignificant contributions of the interactions. Finally, by concatenating $\bsI_a, \forall \bsG_{\bsJ_a}$, one obtains $\bsI \in \mathbb{R}^{\varsigma \times n^2}$.

% But instead of using a specific hyperparameter value $p$,  for each group $\bsG_{\bsJ_a}$ a different value $p_{ab}^k$ is assigned for each pixel $k$ ($k = 1,\cdots, n^2$) and patch $b$ respectively.

\textit{Construct the Hamiltonian kernel:}
In the baseline architecture \dk{of} De-QuIP, for each $\bsJ_a$, the Hamiltonian/energy operator depends on the hyperparameter $(\hbar ^2/2m)$ (\textit{i.e.}, the Planck constant), the total interaction $\bsI_a$ and the original potential/pixels' values $\bsJ_a$. This operator gives the adaptive basis $\bsB_a$ on which the noisy patch $\bsJ_a$ is projected. The integration of the local interactions, bringing a non-local dimension to the formalism, is a core \sd{feature} of De-QuIP.

This physical attribute of the Hamiltonian operator is preserved in this step by constructing a kernel
%%\vspace{-2mm}
\begin{equation}
%%\vspace{-2mm}
\bsC_{2_a} = (\hbar ^2/2m)_a \nabla^2 + \bsJ_a + \bsI_a, ~~\forall \bsJ_a \in \Omega,
\label{eq:hamilt_ker}
\end{equation}
where different learnable values of $(\hbar ^2/2m)_a$ are allotted instead of a constant one. This kernel $\bsC_{2_a}$ mimics the role of the adaptive basis $\bsB_a$ in the next layer in the shadow of a convolutional process. Note that throughout the learning process the kernel retains its original Hamiltonian structure which is a key ingredient of the original De-QuIP algorithm.

\textit{Projection layer:}
This layer deals with the adaptive transformation of the noisy patch $\bsJ_a$ on the associative quantum adaptive basis $\bsB_a$ for each $\bsJ_a \in \Omega$, \textit{i.e.}, $\balpha_a = \bsB_a \bsJ_a$, where $\balpha_a \in \mathbb{R}^{n^2}$ are the projection coefficients of $\bsJ_a$. In our proposed deep architecture, this process is conducted by performing convolution operations on $\bsJ_a$ using a learnable kernel $\bsC_{2_a}$ built in the previous step, as:
%%\vspace{-2mm}
\begin{equation}
%%\vspace{-2mm}
\balpha_a = \bsC_{2_a} \star \bsJ_a, ~~\forall \bsJ_a \in \Omega.
\label{eq:projection}
\end{equation}
Exploiting the power of a deep network, the convolution operation \eqref{eq:projection} removes the algebraically expensive processes, such as the computation of adaptive basis and projection coefficients, and uses the training dataset to directly estimate the projection coefficients. Finally, all $\balpha_a$ are concatenated to form $\balpha \in \mathbb{R}^{\zeta \times n^2}$, serving as input to the next layer.

\textit{Thresholding layer:}
The thresholding layer handles the process of trimming the projection coefficients $\balpha$. A nonlinear ReLU activation function $\varphi$ is used as a thresholding function, which makes the denoising process more robust by adding more flexibility than the baseline scheme, where thresholding was done in energy. Therefore, the shrunk coefficients $\bsR_a = \varphi(\balpha_a)$ are obtained for each $\bsJ_a \in \Omega$, further concatenated into $\bsR \in \mathbb{R}^{\zeta \times n^2}$, before stepping to the next layer.

\textit{Inverse projection layer:}
In the original algorithm the denoised patch $\hat{\bsJ_a}$ is revamped from the reduced coefficients $\bsR_a$ by inverse projecting onto the quantum adaptive basis $\bsB_a$ for each $\bsJ_a \in \Omega$, \textit{i.e.}, $\hat{\bsJ_a} = \bsB_a^{-1} \bsR_a$. This step resembles a convolution process of $\bsR_a$ with a learnable kernel $\bsC_{3_a}$. Hence, the mathematical operation of the layer is defined as
%%\vspace{-2mm}
\begin{equation}
%%\vspace{-2mm}
\hat{\bsJ_a} = \bsC_{3_a} \star \bsR_a, ~~\forall \bsJ_a \in \Omega.
\label{eq:inve_proj}
\end{equation}
Finally, before proceeding to the following layer, all outputs $\hat{\bsJ_a}$ are concatenated to $\hat{\bsJ~} \in \mathbb{R}^{\zeta \times n^2}$.

Note that in the baseline algorithm, the operator used in the inversion step was the inverse of the adaptive basis used in the projection process. This mutual dependence is highlighted in Fig.~\ref{fig:Arch_com}(a) by an arrow. In the proposed deep unfolded network, the learnable kernels $\bsC_{2_a}$ and $\bsC_{3_a}$ replaced respectively the original and inverse adaptive basis. The convolutional operations are useful to learn these kernels independently and are illustrated by removing the arrow in Fig.~\ref{fig:Arch_com}(b).

\textit{Aggregation layer:}
Akin to the De-QuIP scheme, this layer conducts the $\bsE^{-1}$ operation to accumulate all the denoised patches and put them back to their initial positions in the image after normalization, and reconstructs the denoised image $\hat{\bsx}$. Note that overlapping patches are considered in the proposed formalism. Fig.~\ref{subfig:deep_dequip_layers} illustrates the proposed \dk{DIVA} network architecture, highlighting all the layers described above.

%\vspace{-4mm}
\subsection{Proposed DIVA Advanced Network}
\label{sec:deep_dequipa}
%%\vspace{-1mm}
An advanced version of the \dk{DIVA} network introduced in the previous section is proposed hereafter. This network slightly differs from \dk{DIVA}, and is adapted to image restoration tasks involving, in addition to noise, other degradation effects on the observed image $\bsy$, such as blur, pixel resolution loss or missing pixels. In the case of additive Gaussian noise, the effect of the noise and the additional degradation can be considered independently. Therefore, \dk{DIVA} network of Sec.~\ref{sec:deep_dequip} is extended by additional convolutional layers after the inversion process. In this way, the first part of the network eliminates the noise, and the \sd{second} part neutralizes the effects of a nonidentity degradation operator.

The modified network referred to as \dk{DIVA}-A primarily plugs a neutralization layer between the inverse projection and aggregation layers, as highlighted in Fig.~\ref{subfig:deep_dequipA_layers}.

\textit{Neutralization layer:}
This layer corresponds to the restoration of the patch $\tilde{\bsJ_a}$ by eliminating the influence of a degradation operator $\bsO_a$ from the patch $\hat{\bsJ_a}$ reconstructed in the inverse projection layer for each $\bsJ_a \in \Omega$, \textit{i.e.}, $\tilde{\bsJ_a} = \bsO_a^{-1} \hat{\bsJ_a}$, where $\bsO_a$ denotes a degradation operator acting on a patch $\bsJ_a$, $\forall \bsJ_a \in \Omega$. This operation is analogous to a convolutional process of $\hat{\bsJ_a}$ with a learnable kernel $\bsC_{4_a}$, defined as
%%\vspace{-2mm}
\begin{equation}
%%\vspace{-2mm}
\tilde{\bsJ_a} = \bsC_{4_a} \star \hat{\bsJ_a}, ~~\forall \bsJ_a \in \Omega.
\label{eq:neutraliza}
\end{equation}
The proposed network conducts this operation by adding three convolutions with multiple learnable filters, and one ReLU function to remove any unwanted contribution (see Fig.~\ref{subfig:deep_dequipA_layers}). The power of a CNN architecture is used to learn these filters that mimic the role of a degradation operator in this layer.

Before proceeding to the aggregation layer, all $\tilde{\bsJ_a}$ are concatenated to obtain $\tilde{\bsJ~} \in \mathbb{R}^{\zeta \times n^2}$. Similar to the \dk{DIVA} network, the aggregation layer assembles all recovered patches and outputs the restored image $\hat{\bsx}$.

%\vspace{-4.5mm}
\subsection{Loss Function}
\label{sec:loss_func}
%%\vspace{-1mm}
The proposed networks are trained end-to-end, where the mean squared error (MSE) between the predicted and original residuals is adopted as the loss function \cite{Yang2018bm3dnet}:
%%\vspace{-2mm}
\begin{equation}
%%\vspace{-2mm}
\mathcal{L}_\Theta = \dfrac{1}{MN} \norm { \mathcal{R}(\hat{\bsx}; \Theta)  - (\bsy - \bsx) }^2_2,
\label{eq:loss_fun}
\end{equation}
where $\mathcal{R}(\hat{\bsx}; \Theta)$ denotes the predicted residual by the network with parameter set $\Theta$. This loss function allows our models to learn the disorders present in a distorted image without bothering about the features of the true image.
\dk{Note that it is possible to use different other loss functions.}
\footnote{The Python code of the proposed the trained networks are available at \href{https://github.com/SayantanDutta95/}{github.com/SayantanDutta95/}}

%\vspace{-4.5mm}
\section{Experimental Results}
\label{sec:expe_results}
%%\vspace{-1.5mm}
In this section, we analyze the proposed networks and illustrate their performance in various image restoration tasks, such as image denoising, deblurring, SR, and inpainting.

Sec.~\ref{sec:exp_setting} briefly summarizes the experimental settings used in the different contexts. Sec.~\ref{sec:comp_meth} gives an overview of various benchmark methods considered for comparison purposes. An ablation study with/without considering the interaction layer and the Hamiltonian kernel within the proposed networks is conducted in Sec.~\ref{sec:ablation}, with an additional discussion on the parameter number, run time, and the depth of the network. Finally, Sec.~\ref{sec:results} presents a quantitative and qualitative evaluation of our DL models on various image restoration problems.

%\vspace{-4mm}
\subsection{Experimental Settings}
\label{sec:exp_setting}

\subsubsection{Image Denoising}
%%\vspace{-1mm}
\textit{Training data.}
The proposed \dk{DIVA} network was trained for the Gaussian denoising task following \cite{Chen2017trainable, Zhang2017beyond, Zhang2018FFDNet}, over a set of $400$ gray-scale images of size $180 \times 180$ extracted from BSD400 dataset. All images were contaminated with additive white Gaussian noise (AWGN) with standard deviation $\sigma$, following two configurations: known and unknown $\sigma$. For the case of known $\sigma$, the training was conducted individually over six known noise levels, for $\sigma = 10, 15, 25, 50, 75$ and $100$. To tackle an unknown noise level, \dk{DIVA} was also trained blindly for a range of noise levels corresponding to $\sigma \in [5,40]$. The corresponding model is referred as \dk{DIVA}-blind.
 
\textit{Testing data:}
The trained networks were tested on five standard benchmark datasets Set12, BSD68, Kodak, LIVE1 and Urban100,
% containing 12, 68, 25, 30 and 100 images respectively
widely-used for denoising problems \cite{Zhang2017beyond, Zhang2018FFDNet}.

%\vspace{-2mm}
\subsubsection{Image Deblurring}
%%\vspace{-1mm}
\textit{Training data.}
\dk{DIVA}-A was trained separately for two types of blur kernels, \textit{i.e.}, motion and Gaussian blur, using the recently released high-quality dataset DIV2K \cite{agustsson2017ntire} that consists of 800 images. Eight real motion blur (MB) kernels \cite{Levin2011efficient, Kong2022deep} and three Gaussian blur (GB) kernels \cite{Wang2018training} were considered with AWGN. %We conducted the training of each model for different blur settings.

\textit{Testing data:}
The models trained for motion blur were tested on four benchmark datasets Set10, Levin, Sun \textit{et al.}, and Set12, used in \cite{Kong2022deep, Nan2020variational}. The BSD100 and Set16 datasets were considered for the Gaussian case, following \cite{Wang2018training}.

\subsubsection{Single Image Super-Resolution}

\textit{Training data.}
Similar to the deblurring model, the high-quality DIV2K \cite{agustsson2017ntire} dataset was used as training data for image SR application.
Two degradation models were used to simulate LR images for network training: (i) bicubic downsampling (BD), and (ii) Gaussian downsampling (GD). The scaling factor was set to x2, x3, and x4. For BD case \cite{Anwar2022densely}, a LR image was simulated from the HR image by adopting Matlab \textit{imresize} function, whereas for GD scenario, the HR image was blurred by a Gaussian kernel of size $7 \times 7$ with standard deviation $1.6$ before downsampling, similar to \cite{Anwar2022densely}.

\textit{Testing data:}
For testing, four widely-used benchmark datasets for image SR problem \cite{Anwar2022densely, Ahn2018fast, Tai2017MemNet} Set5, Set14, BSD100, and Urban100, were used.

\subsubsection{Image Inpainting}
%%\vspace{-1mm}
\textit{Training data:}
The same 400 gray-scale images \cite{Zhang2017beyond} exploited by the denoising model were used to conduct the training of the proposed \dk{DIVA}-A model for image inpainting. Random pixel missing model was considered to generate LR images from HR ones. 20\%, 50\% and 80\% rates of missing pixels were used.

\textit{Testing data:}
Datasets Set5 and Set12 were used to evaluate the trained inpaining networks.

\subsubsection{Quantitative Metrics}

For the purpose of quantitative evalution, the peak-signal-to-noise-ratio (PSNR) and the structural similarity (SSIM)
%\cite{wang2004image}
computed between the true and the restored images were used.

\subsubsection{Training Settings}

All HR and simulated LR images were clipped between $0$ and $1$. The patch size was set to $n = 15$ with a local window of size $W = 35$ for the proposed image denoising model with known $\sigma$. For \sd{DIVA-blind and inpainting applications,} these parameters were slightly modified to $n = 25$ and $W = 50$. For deblurring and SR, larger patch and window sizes were used, $n = 35$ and $W = 70$, to preserve more spatial information from the local neighborhood.  
Finally, all LR-HR patch pairs were augmented randomly by rotating 90 degree and flipping horizontally or vertically to generate training data pairs. The proposed models were trained in a supervised manner by exploiting these patch-pairs.

%\textit{i.e.}, in the case where $\sigma$ is unknown, and for the application to inpainting,

To conduct the training, the ADAM optimizer with a mini-batch size of $128$ was employed. More precisely, the models were trained with an exponentially decaying learning rate ranging from $10^{-3}$ to $10^{-6}$ over $60$ epochs. The proposed network architectures were implemented under the Keras framework, and trained using NVIDIA GTX 1080 Ti GPU. The training process took about 6 hours for DIVA and 12 hours for DIVA-A to reach convergence for each experiment.

% of the proposed DL models
% the Keras and Keras-backend framework that relies on the TensorFlow library, /  less than

%%\vspace{-5mm}
\subsection{Comparison Methods}
\label{sec:comp_meth}
%%\vspace{-1mm}
This subsection regroups the state-of-the-art methods used to conduct a comprehensive comparison to illustrate the potential of the proposed models in various imaging problems. 

%\vspace{-2mm}
\subsubsection{Image Denoising}

The residual learning-based DnCNN \cite{Zhang2017beyond} model is the benchmark for AWGN denoising, and its superiority over model-based (e.g., BM3D \cite{Dabov2007Image}, NLM \cite{buades2005review}%buades2008nonlocal
, etc.), and learning-based (e.g., TNRD \cite{Chen2017trainable}, MLP \cite{Burger2012image}, CSF \cite{schmidt2014shrinkage} etc.) algorithms is well-established. In addition to DnCNN \cite{Zhang2017beyond}, our denoising model \dk{DIVA} was also compared \dk{to} two recently introduced DL-based networks, FFDNet \cite{Zhang2018FFDNet} and IRCNN \cite{Zhang2017learning}. Furthermore, comparisons were carried out with a newly proposed deep unfolded scheme, BM3D-NET \cite{Yang2018bm3dnet}, as well as with the baseline De-QuIP \cite{dutta2022novel} algorithm. 

% denoising

%\vspace{-2mm}
\subsubsection{Image Deblurring and SR}

For image deblurring and SR problems, newly published leading methods from the literature were considered to illustrate the accuracy of \dk{DIVA}-A architecture. In the following, the relevant methods used for comparison purposes in different settings are listed.
(i) MB model: IDD-BM3D\cite{Danielyan2012bm3d}, FDN\cite{Kruse2017learning}, VEMNet\cite{Nan2020variational}, DWDN\cite{Dong2020deep}, DRED-DUN\cite{Kong2022deep};
(ii) GB model: IDD-BM3D\cite{Danielyan2012bm3d}, Son \textit{et al.}\cite{Son2017fast}, DEBCNN\cite{Wang2018training};
(iii) BD model: LapSRN\cite{Lai2017deep}, MemNet\cite{Tai2017MemNet}, CARN\cite{Ahn2018fast}, DRLN\cite{Anwar2022densely};
(iv) GD model: IRCNN\cite{Zhang2017learning}, DFAN\cite{Li2022DFAN}, RDN\cite{Zhang2021residual}, DRLN\cite{Anwar2022densely}.

In image SR problems, the DRLN \cite{Anwar2022densely} is the new benchmark in the literature. It is already shown in the seminal paper that the DRLN \cite{Anwar2022densely} exhibits \sd{reference state-of-the-art performance for image SR.}
% % %\sd{against \cite{Dong2016image, Dong2016Accelerating, Wang2015deep, Kim2016Accurate, Lim2017Enhanced}.}
%against the SRCNN \cite{Dong2016image}, FSRCNN \cite{Dong2016Accelerating}, SCN \cite{Wang2015deep}, VDSR \cite{Kim2016Accurate}, EDSR \cite{Lim2017Enhanced}, RCAN \cite{Zhang2018image}.
Thus, the DRLN \cite{Anwar2022densely} was considered in the comparisons, thus avoiding to include all the other approaches. Similarly, for image deblurring, DWDN \cite{Dong2020deep}, DRED-DUN\cite{Kong2022deep}, and DEBCNN\cite{Wang2018training} were the best performing models in their fields. Hence, these models are selected for comparisons over other methods in the literature.
%, such as DPDNN \cite{Dong2019denoising}, RGDN \cite{Gong2020learning}, DCNN \cite{Xu2014deep} or USRNet \cite{zhang2020deep}.\dk{REMOVE THESE 4 LAST METHODS since you didn't use them}

%\vspace{-2mm}
\subsubsection{Image Inpainting}

\dk{DIVA}-A trained for image inpainting was compared against the DL prior based model IRCNN \cite{Zhang2017learning}.

The pretrained models and the testing codes, made publicly available by the authors, were used for comparisons. Importantly, note that the proposed networks have been trained and tested exactly in the same conditions and on the same datasets as the comparison methods, thus ensuring a fair comparison.

%\vspace{-4mm}
\subsection{Ablation Study and Model Analysis}
\label{sec:ablation}

This section regroups several ablation studies aiming at showing the importance of the layers inspired from quantum mechanics, and an in-depth analysis of the properties of the proposed networks.

%\vspace{-2mm}
\subsubsection{Influence of the Interaction Layer}

To show the effect of the interaction layer's integration in the Hamiltonian kernel, two versions of the \dk{DIVA} network were trained for image denoising with $\sigma = 15$: the complete network as shown in Fig. \ref{fig:deep_model_layers}(a), and the same network without the interaction layer. Fig.~\ref{fig:ablation_loss_func} plots the corresponding loss functions for these two network settings with respect to the number of epochs. One can see that using the interaction layer results into a faster and more stable convergence of the training process. Meanwhile, in the absence of this layer, a strong periodic fluctuation can be observed. This is caused by the absence of a non-local architecture in the network, which helps stabilizing the convergence process.

The same ablation study was conducted for different depths of the projection layer, using the Hamiltonian convolutional kernel constructed with and without the interaction layer. From Table~\ref{tab:tab_ablation_depth_interaction}, one can see a clear improvement in denoising performance in the presence of the interaction layer. In addition, the interaction layer significantly reduces the depth of the network by extracting the local similarities/structures from the neighboring patches. Indeed, more local information can be transferred through this non-local architecture, which helps network structures with lower depth to be more efficient. On the contrary, the network without the interaction layer improves while increasing the depth. This is expected since a deeper network consists of a larger set of tunable parameters. Although a bigger set of parameters leads to a better outcome, the learning process becomes more computationally expensive. Thus, the integration of the interaction layer enhances the network performance with a reduced computational cost, giving an edge to the proposed models.

% add figures  --------------------------------------
\begin{figure}[t!]
\begin{centering}

\includegraphics[width=0.39\textwidth]{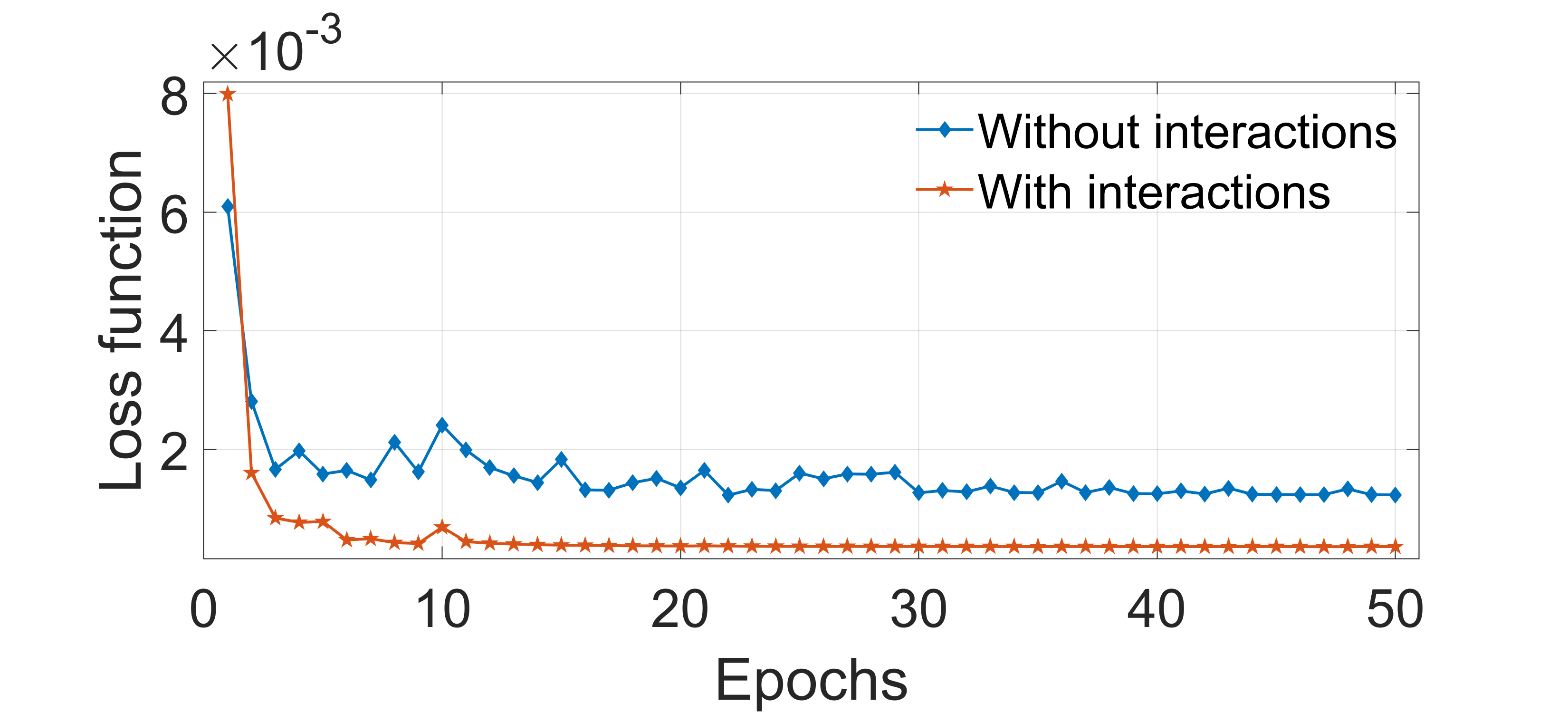}

\end{centering}
%\vspace{-2mm}
\caption{Loss function (MSE) with respect to epochs.
Two specific models are trained for image denoising with $\sigma = 15$, with and without integrating the interaction layer in the proposed DIVA architecture for the ablation study.}
%\vspace{-5mm}
\label{fig:ablation_loss_func}
\end{figure}

\begin{table*}[t!]
\setlength\tabcolsep{4pt}
\begin{scriptsize}
\begin{center}
\caption{Ablation investigation of the projection layer's depth using Hamiltonian kernel with or without the interaction layer. The results (PSNR/SSIM) are obtained on Set12 contaminated with AWGN with $\sigma = 15$, in 50 epochs.}
%%\vspace{-4mm}
\label{tab:tab_ablation_depth_interaction}
\begin{tabular}{c ccccc c ccccc}

\hline
 & \multicolumn{11}{c}{Depth of the projection layer using Hamiltonian convolutional kernel}\\ 

& 1 & 2 & 3 & 4 & 5 && 1 & 2 & 3 & 4 & 5 \\
\hline
%\cline{2-9}
%---------------------------------------------------

Interaction layer    & \xmark & \xmark & \xmark & \xmark & \xmark && \cmark & \cmark & \cmark  & \cmark & \cmark \\

\hline

PSNR(dB)/SSIM(\%)    & 30.38/87.64 & 31.61/89.22 & 31.95/90.74 & 32.17/91.61 & 32.28/91.88 && 32.09/93.68 & 32.92/95.41 & 32.95/95.52 & 32.96/95.55 & 32.98/95.60 \\

\hline

\end{tabular}\end{center}
\end{scriptsize}
%\vspace{-4mm}
\end{table*}

\begin{table}[t!]
\setlength\tabcolsep{10pt}
\begin{scriptsize}
\begin{center}
\caption{Ablation study with/without using the Hamiltonian kernel in the network. The results (PSNR/SSIM) are obtained in 50 epochs on Set12 images contaminated with AWGN ($\sigma = 15$).}
%%\vspace{-4mm}
\label{tab:tab_ablation_Hamiltonian}
\begin{tabular}{l ccc}
\hline
 & \multicolumn{3}{c}{Contribution of different components}\\ 
\hline
%\cline{2-9}
%---------------------------------------------------

Hamiltonian kernel	& \xmark & \cmark & \cmark \\

Interaction layer	& \xmark & \xmark & \cmark \\
\hline

PSNR(dB)/SSIM(\%)	& 29.30/86.82 & 31.61/89.22 & 32.92/95.41 \\

\hline
\end{tabular}\end{center}
\end{scriptsize}
%\vspace{-4mm}
\end{table}

% add figures  --------------------------------------
\begin{figure*}[t!]
\begin{centering}

\subfigure[Left: Denoising performance vs parameter number comparisons are presented on the BSD68 dataset with $\sigma=50$.
Right: Denoising performance vs run time comparisons are presented on the Set12 dataset with $\sigma=50$.] {\label{subfig:Time_Parameter_denoi}
\includegraphics[width=.32\textwidth]{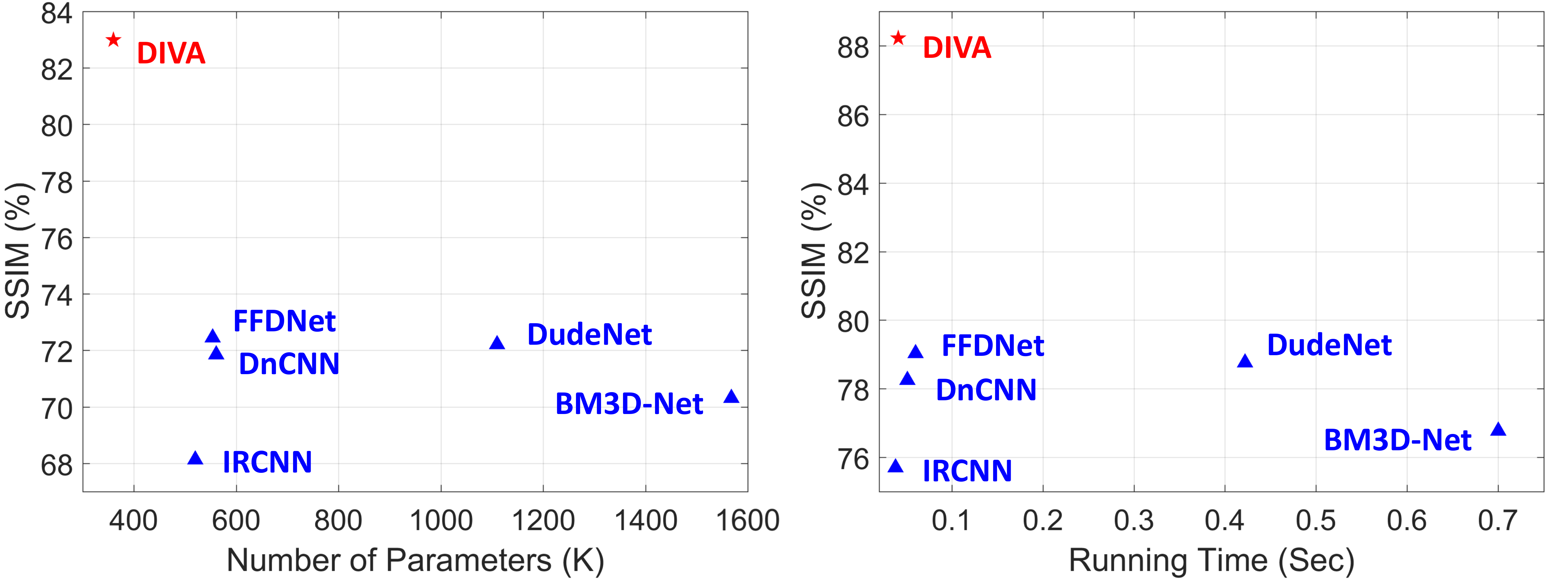}}
\subfigure[Left: Deblurring performance vs parameter number comparisons are presented on the Levin dataset with motion blur and $\sigma=7.65$.
Right: Deblurring performance vs run time comparisons presented on the Levin dataset with motion blur and $\sigma=7.65$.] {\label{subfig:Time_Parameter_deblur}
\includegraphics[width=.32\textwidth]{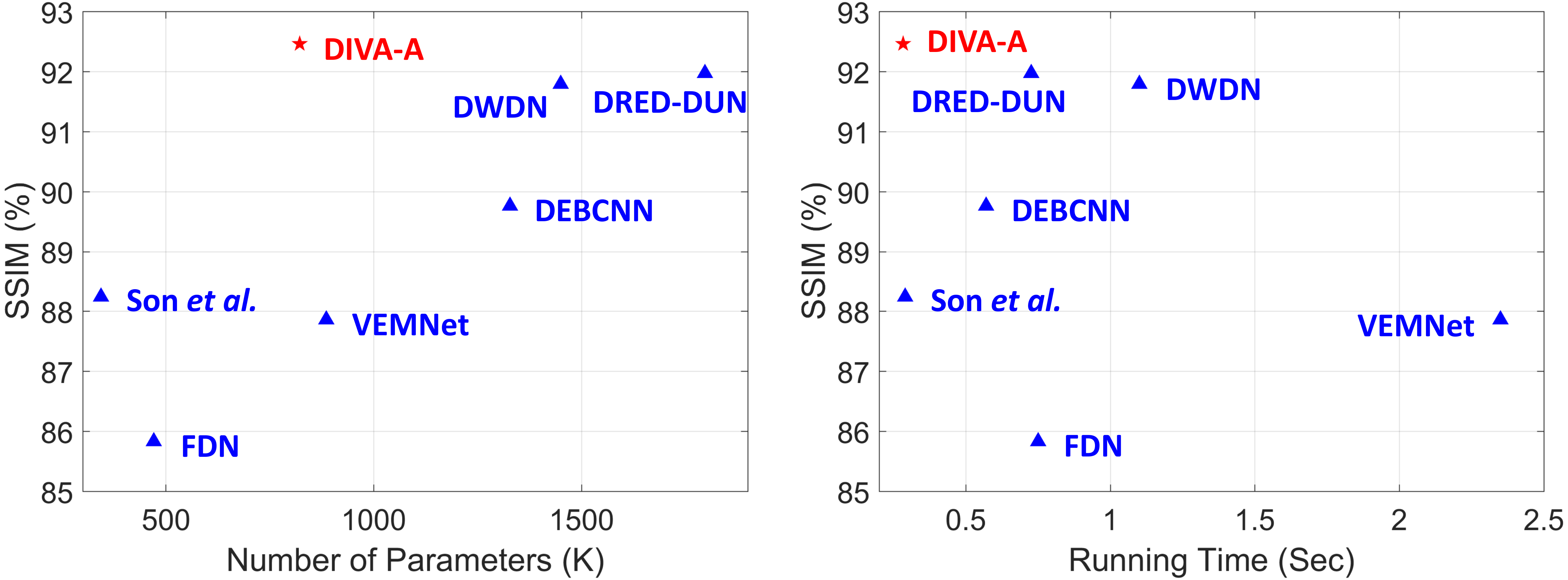}}
\subfigure[Left: SR performance vs parameter number comparisons are presented on the BSD100 dataset for 4X SR.
Right: SR performance vs run time comparisons are presented on the Urban100 dataset for 4X SR.] {\label{subfig:Time_Parameter_sr}
\includegraphics[width=.32\textwidth]{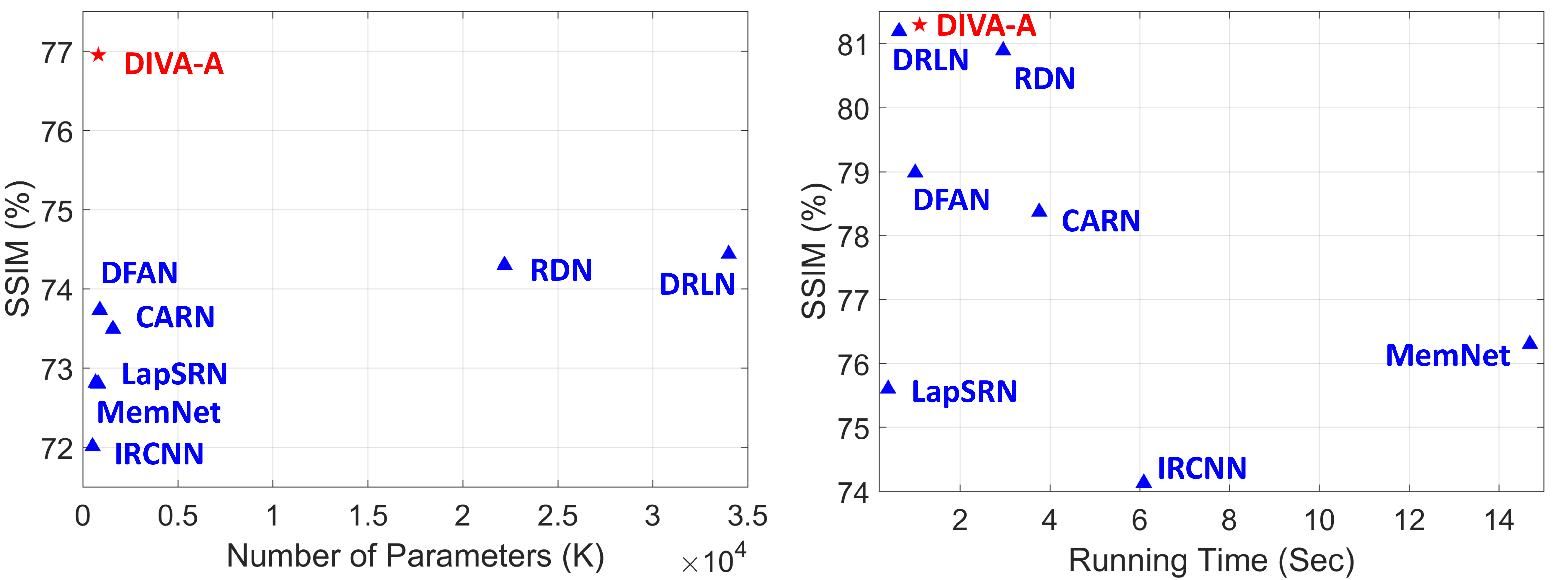}}

\end{centering}
%\vspace{-2mm}
\caption{Performance versus parameter number and run time versus performance are presented for different methods for different tasks. The proposed methods give high performances in terms of SSIM(\%) with fewer number of parameters and low computation time.}
%\vspace{-5mm}
\label{fig:Time_Parameter_SSIM}
\end{figure*}

Note that, in absence of the interaction layer in the proposed model, the network does not consider the influence of neighboring patches on the target patch and loses its non-local nature. Thus, each patch behaves as a single particle quantum system, and all patches are independent. Hence, in this circumstance, the network without an interaction layer becomes an unfolded DL scheme of the baseline QAB algorithm \cite{dutta2021quantum}, originally proposed for image denoising based on single-particle quantum theory. Recently, in \cite{dutta2022novel}, it has been shown that the baseline De-QuIP outperforms the conventional QAB algorithm significantly. This observation by the traditional algorithm \cite{dutta2022novel} is also consistent in our unfolded DL models, as reported in Fig.~\ref{fig:ablation_loss_func} and Table~\ref{tab:tab_ablation_depth_interaction}. Therefore, the consideration of the quantum interaction concept \sd{clearly} enhances the model performance of both conventional and DL architectures.

% profoundly

% The results reported by our unfolded DL models in the Fig.~\ref{fig:ablation_loss_func} and Table~\ref{tab:tab_ablation_depth_interaction} are consistent with the observations made by the traditional algorithms \cite{dutta2021image, dutta2022novel}.

%\vspace{-2mm}
\subsubsection{Depth of the Projection Layer}

% Ablation study on the depth of the projection layer

Table~\ref{tab:tab_ablation_depth_interaction} reports denoising performance on Set12 for $\sigma = 15$ for different depths of the projection layer within the Hamiltonian kernel. As expected, the denoising performance increases with the depth of the network, but this increment is less significant beyond depth $3$. Assessing the trade-off between the network efficiency and the computational complexity, a depth of $2$ was considered in the proposed DL models.

%\vspace{-2mm}
\subsubsection{Ablation Study on the Hamiltonian Kernel}
%%\vspace{-1mm}

In the proposed models, the objective is to construct a Hamiltonian kernel to conduct the projection operation, while preserving the original attributes of the proposed Hamiltonian operator in the baseline De-QuIP algorithm \cite{dutta2022novel}. This Hamiltonian kernel is a sum of the nabla operator, original pixels' values of the patch and the interactions with its neighbors, following equation \eqref{eq:hamilt_ker}.
To illustrate the importance of this Hamiltonian structure in the proposed networks, an ablation investigation of this Hamiltonian kernel was conducted, through three network settings: (i) without the Hamiltonian kernel and interaction layer,
(ii) with the Hamiltonian kernel but without the interaction layer, and (iii) with the Hamiltonian kernel including the interaction layer. For all settings, the depth of the projection layer was set to 2. Table~\ref{tab:tab_ablation_Hamiltonian} regroups the denoising results on Set12 for AWGN with $\sigma = 15$ for all these three configurations. From these results, one may observe that the accuracy of the network is significantly improved in the case where the Hamiltonian shape is preserved and includes the interactions between neighboring patches. This improvement is even further illustrated by the SSIM, that is more sensitive to the image structure than the PSNR, and thus more suitable to reflect the contribution of the interaction-based Hamiltonian operator. 
% Inclusion of the Hamiltonian kernel with the interaction layer in the network increases performance, while the potential of the network drops when the interaction layer is absent. 
Furthermore, one may notice that without none of these two ingredients, the denoising performance is largely decreased. This can be explained by the fact that in this case, the resulting netwrok, very similar to DnCNN \cite{Zhang2017beyond}, needs far more layers to achieve good denoising results. Indeed, a network depth of $17$ is suggested in \cite{Zhang2017beyond}, while, as mentionned \dk{previously}, the proposed network depth can be reduced to $2$.
% because the depth of the network is not optimal if the Hamiltonian kernel and the interaction layer are missing in the network. Indeed, when these two settings are absent, our proposed \dk{DIVA} network resembles the DnCNN \cite{DnCNN} network, and in that seminal paper, the optimal depth is set to 17 for image denoising. Thus, the drop in network performance in this regard is understandable and justifiable. 
Therefore, the exploitation of the local information through the \dk{patch} interaction, originally proposed in the baseline De-QuIP, and the attributes of the Hamiltonian kernel, make the proposed DL networks easily adaptable but resilient even for lower depth. 
In conclusion, this experiment illustrates the significance of the inclusion of the Hamiltonian kernel with the interaction layer in the proposed models.

% The performance is higher when the Hamiltonian kernel with the interaction layer is included in the network, while the potential of the network drops when the interaction layer is absent.

%\vspace{-2mm}
\subsubsection{Analysis of the Parameter Number and Runtime}
%%\vspace{-1mm}

The number of hyperparamers of a DL network plays a crucial role in its efficiency. Generally, a larger pool of parameters drives the model more resilient and leads to better performance. However, it also imposes an important computational load, in particular within the training process. Furthermore, excess baggage of parameters may lead to an over-fitting problem. Hence, a balanced trade-off between the learnable parameter number, the performance, and the computational cost becomes a crucial factor for an efficient DL model.

\begin{table*}[t!]

\begin{scriptsize}
\begin{center}
\caption{Image denoising results in terms of average PSNR(dB) and SSIM(\%) values for five benchmark datasets contaminated by six noise levels ($\sigma=10,15,25,50,75,100)$. For each experiment, the best values are in bold and the second best values are underlined.}
%%\vspace{-4mm}
\label{tab:tab_psnr_ssim_image_denoising}
\begin{tabular}{l c c ccccccc}
%%%%%%%%%%%%%%%%%%%%%%%%%Noise = 20%%%%%%%%%%%%%%%%%%%%%%%%%%%
\hline

%\multirow{2}{*}{Methods} & BSD68 & Kodak & Urban100 & LIVE1 & Set12 \\ \cline{2-5}

Dataset & $\sigma$ & Input & \multicolumn{7}{c}{Methods}\\ 

& & & DnCNN\cite{Zhang2017beyond} & FFDNet\cite{Zhang2018FFDNet} & IRCNN\cite{Zhang2017learning} & BM3D-NET\cite{Yang2018bm3dnet} & De-QuIP\cite{dutta2022novel} & DIVA & DIVA-blind\\
\hline
%\cline{2-8}

%---------------------------------------------------

\multirow{6}{*}{Set12}

& 10  & 28.16/82.87 & \bdb{34.76}/92.69 & 34.64/92.71 & 33.62/91.83 & 33.27/91.97 & 33.45/91.03 & \bdr{34.80/96.77} & 34.68/\bdb{94.56}\\

& 15  & 24.64/69.97 & \bdb{32.84}/90.23 & 32.75/90.27 & 32.77/88.08 & 31.65/88.96 & 31.15/87.30 & \bdr{32.92/95.41} & 32.79/\bdb{93.61}\\

& 25  & 20.20/49.68 & \bdb{30.42}/86.14 & \bdb{30.42}/86.34 & 30.38/84.23 & 29.77/85.09 & 28.65/81.23 & \bdr{30.47/93.00} & 30.36/\bdb{90.73}\\

& 50  & 14.18/24.87 & 27.16/78.25 & \bdb{27.32/79.03} & 27.14/75.70 & 25.78/76.77 & 25.28/70.43 & \bdr{27.45/88.22} & -/- \\

& 75  & 10.66/14.75 & 25.15/71.71 & \bdb{25.49/73.52} & 23.75/67.46 & -/- & 23.44/63.69 & \bdr{25.63/84.31} & -/- \\

& 100 &  8.16/9.64  & 23.87/64.28 & \bdb{24.20/69.26} & 21.95/59.70 & -/- & 22.21/58.02 & \bdr{24.43/81.17} & -/- \\

%\cline{2-8}
\hline

%---------------------------------------------------

\multirow{6}{*}{BSD68}

& 10  & 28.15/83.57 & \bdb{33.87}/92.71 & 33.75/92.66 & 33.74/90.57 & 32.74/91.73 & 32.67/90.65 & \bdr{33.94/96.21} & 33.80/\bdb{94.38}\\

& 15  & 24.63/70.99 & \bdb{31.73}/89.06 & 31.63/89.02 & 31.63/87.98 & 31.42/88.77 & 30.24/85.38 & \bdr{31.79/94.04} & 31.64/\bdb{92.74}\\

& 25  & 20.19/50.70 & \bdb{29.22}/82.78 & 29.19/82.89 & 29.15/79.51 & 28.95/81.42 & 27.83/77.35 & \bdr{29.34/90.07} & 29.19/\bdb{87.44}\\

& 50  & 14.17/25.08 & 26.22/71.85 & \bdb{26.29/72.45} & 26.16/68.13 & 25.73/70.31 & 24.88/64.25 & \bdr{26.33/82.99} & -/- \\

& 75  & 10.65/14.61 & 24.63/64.69 & \bdb{24.78/65.86} & 22.87/60.05 & -/- & 23.33/56.55 & \bdr{24.87/77.81} & -/- \\

& 100 & 8.15/9.41   & 23.16/55.46 & \bdb{23.77/60.96} & 19.46/49.47 & -/- & 22.27/51.23 & \bdr{23.93/74.21} & -/- \\

%\cline{2-8}
\hline

%---------------------------------------------------

\multirow{6}{*}{Kodak}

& 10  & 28.14/81.24 & \bdb{34.86}/92.17 & 34.81/92.20 & 34.76/87.91 & 32.39/91.01 & 33.56/89.95 & \bdr{34.91/96.35} & 34.82/\bdb{94.75}\\

& 15  & 24.62/67.32 & \bdb{32.84}/88.82 & 32.72/88.90 & 32.63/83.40 & 30.82/87.68 & 31.27/85.13 & \bdr{32.93/94.49} & 32.78/\bdb{93.02}\\

& 25  & 20.18/45.89 & \bdb{30.43}/83.15 & 30.37/83.42 & 30.29/78.07 & 28.55/81.62 & 28.83/77.64 & \bdr{30.55/91.16} & 30.30/\bdb{87.89}\\

& 50  & 14.16/21.13 & 27.47/73.53 & \bdb{27.61/74.34} & 27.44/69.24 & 25.91/72.15 & 25.71/65.76 & \bdr{27.70/85.41} & -/- \\

& 75  & 10.64/11.91 & 25.77/67.34 & \bdb{25.96/68.80} & 23.85/61.75 & -/- & 24.07/59.02 & \bdr{26.16/81.36} & -/- \\

& 100 & 8.14/7.54 & 23.99/55.99 & \bdb{24.88/64.74} & 20.38/51.29 & -/- & 22.92/53.74 & \bdr{25.22/78.66} & -/- \\

%\cline{2-8}
\hline

%---------------------------------------------------

\multirow{6}{*}{LIVE1}

& 10  & 28.14/83.19 & \bdb{34.24}/92.95 & 34.13/92.96 & 33.02/88.09 & 32.77/91.83 & 32.39/90.98 & \bdr{34.27/96.54} & 32.19/\bdb{94.81}\\

& 15  & 24.62/70.46 & \bdb{32.11}/89.68 & 32.01/89.71 & 30.42/81.32 & 30.46/88.74 & 29.96/85.96 & \bdr{32.19/94.65} & 31.97/\bdb{92.69}\\

& 25  & 20.18/50.19 & \bdb{29.55}/83.91 & 29.53/84.08 & 27.22/75.04 & 27.61/82.14 & 27.44/78.00 & \bdr{29.62/91.12} & 29.46/\bdb{88.32}\\

& 50  & 14.16/25.03 & 26.40/73.34 & \bdb{26.51/74.03} & 23.05/66.92 & 24.75/71.60 & 24.28/64.73 & \bdr{26.63/84.54} & -/- \\

& 75  & 10.64/14.74 & 24.70/66.14 & \bdb{24.92/67.59} & 21.21/57.58 & -/- & 22.62/56.66 & \bdr{24.99/79.65} & -/- \\

& 100 & 8.14/9.59 & 22.39/50.10 & \bdb{23.81/62.74} & 19.59/48.28 & -/- & 21.51/50.87 & \bdr{23.99/76.23} & -/- \\

%\cline{2-8}
\hline

%---------------------------------------------------

\multirow{6}{*}{Urban100}

& 10  & 28.15/87.17 & 34.43/95.74 & 34.45/94.89 & 32.93/91.35 & 32.53/94.52 & 31.25/93.26 & \bdr{34.75/97.84} & \bdb{34.52/95.37}\\

& 15  & 24.63/77.13 & 32.17/93.36 & \bdb{32.42}/92.73 & 30.30/88.77 & 30.65/91.99 & 29.53/88.60 & \bdr{32.51/96.52} & 32.26/\bdb{94.11}\\

& 25  & 20.19/60.04 & 29.27/88.42 & \bdb{29.92}/88.87 & 27.01/83.09 & 27.68/86.63 & 25.75/82.53 & \bdr{30.01/93.73} & 29.75/\bdb{91.89}\\

& 50  & 14.17/34.98 & 25.46/77.82 & \bdb{26.52/80.57} & 22.79/71.51 & 23.99/75.34 & 22.81/68.02 & \bdr{26.67/87.80} & -/- \\

& 75  & 10.65/22.46 & 23.23/68.69 & \bdb{24.52/73.65} & 20.81/61.21 & -/- & 20.59/58.92 & \bdr{24.80/82.10} & -/- \\

& 100 & 8.15/15.38 & 22.04/62.85 & \bdb{23.08/67.59} & 18.79/53.57 & -/- & 19.65/50.51 & \bdr{23.39/77.37} & -/- \\

%\cline{2-8}
\hline

%---------------------------------------------------

\end{tabular}\end{center}
\end{scriptsize}
%\textit{$\ast$The best and second best results are highlighted in red and blue, respectively.}
%\vspace{-5mm}
\end{table*}

As detailed in the previous ablation studies, the proposed models exploit the Hamiltonian kernel, which is enriched by an intrinsic non-local architecture through the interaction layer. As a result, the resulting DL networks are able to process more information through fewer parameters and significantly reduce the cost of training with high efficiency. Fig.~\ref{fig:Time_Parameter_SSIM} provides the performance in terms of SSIM(\%) versus the number of parameters and the runtime of the proposed models against state-of-the-art methods, in the context of different image restoration problems. One can observe a significant gain in performance of \dk{DIVA} model for image denoising (see Fig.~\ref{subfig:Time_Parameter_denoi}). \dk{DIVA} increases SSIM by 10\%, with almost half the number of parameters of its closest competitors FFDNet \cite{Zhang2018FFDNet} and DnCNN \cite{Zhang2017beyond}. For image deblurring problem (see Fig.~\ref{subfig:Time_Parameter_deblur}), \dk{DIVA}-A requires only half of the parameters compared to its nearest rival DRED-DUN \cite{Kong2022deep}, but offers a 1\% better SSIM value. Similarly, from Fig.~\ref{subfig:Time_Parameter_sr}, one can report a gain of 1-2\% in SSIM for image SR by \dk{DIVA}-A compared to the recently introduced DRLN network, whereas our model has $40$ times less parameters than DRLN. Naturally, the proposed networks that need a reduced number of parameters to perform well, also offer a signaficantly reduced training cost. Fig.~\ref{fig:Time_Parameter_SSIM} presents the runtime comparisons against other standard models in various imaging tasks, showing that the proposed models are significantly faster. Note that similar results are achieved for image inpainting, but are not reported here since the comparison network is IRCNN, already included in the SR experiments.
Hence, harnessing the benefits of the interaction layer and of the Hamiltonian kernel, the proposed DL models demonstrate better performance for image restoration with fewer parameters and more efficient computational costs.

% The results are consistant with the findings in \cite{dequip}.

%\vspace{-4mm}
\subsection{Qualitative and Quantitative Image Restoration Results}
\label{sec:results}

\subsubsection{Image Denoising}
\label{sec:result_denoi}
%%\vspace{-1mm}

% The efficiency of the proposed algorithm is evaluated in the scenario of additive Gaussian noise.

Table~\ref{tab:tab_psnr_ssim_image_denoising} summarizes the average PSNR and SSIM results of the different methods on six commonly used testing datasets with six different noise levels. One might notice that the proposed \dk{DIVA} model uniformly outperforms all the state-of-the-art approaches, irrespective of the noise level and dataset. Compared to the deep unfolded BM3D network BM3D-NET, our model exhibits much better denoising performance with an average increment of 1.5dB PSNR and 4.5\% SSIM for low noise levels and up to 2dB PSNR and 13\% SSIM for higher $\sigma$. Note also that BM3D-NET was only available for four levels of noise. One can observe that the performance gain is much higher over the benchmark DnCNN and FFDNet networks for high noise cases. Precisely, \dk{DIVA} outperforms these competing methods by 0.05-1.2dB PSNR and 4-18\% SSIM in average and achieves the best denoising yields. Moreover, our blind denoising model \dk{DIVA}-blind that, in contrast to the other networks, is not trained for a given (known) noise level, but for a range of $\sigma$, still gives comparable PSNR values and improved SSIM compared to the state-of-the-art approaches. In all the cases, one can observe a considerable improvement in SSIM enabled by the proposed network, which proves that it is better equipped for image structure and pattern preservation than other models.
%\sd{This observation is consistent with our model architecture. Indeed, our proposed model is powered by a quantum patch interaction formalism, originally proposed in the baseline De-QuIP algorithm, that brings an intrinsic non-local architecture to the network. Exploiting this quantum interaction theory, our network efficiently preserves the image structures/features from a local image neighborhood compared to other standard models.}
\sd{Utilization} of this local information from neighboring patches enables our network to be resilient and adapted to high and low-\dk{level} noise, giving us an edge over other models.
%%\vspace{-4.3mm}

% Absorption

\sd{Furthermore, a notable gain of an average of 1.5-3dB PSNR and 5-26\% SSIM is observed compared to the baseline De-QuIP method. This is a consequence of finely tuned hyperparameters values for each patch by harnessing the power of the backpropagation architecture.}

%%\vspace{-4.3mm}

%Furthermore, a notable gain of an average of 1.5-3dB PSNR and 5-26\% SSIM is observed compared to the baseline De-QuIP method. This is a consequence of properly tuned hyperparameters values by harnessing the power of the CNN network. As discussed in the Subsection~\ref{sec:short_dequip}, using a fixed hyperparameter value for all patches in the system is a major limitation of the original De-QuIP algorithm which limits the adaptability of the model. The deep unfolded network solves this challenge by assigning a correctly tuned unique hyperparameter for each patch, exercising the back propagation architecture. %Thus, \dk{DIVA} model shows significantly better performance over baseline De-QuIP.

% add figures  --------------------------------------
\begin{figure*}[t!]
\begin{centering}

\includegraphics[width=1\textwidth]{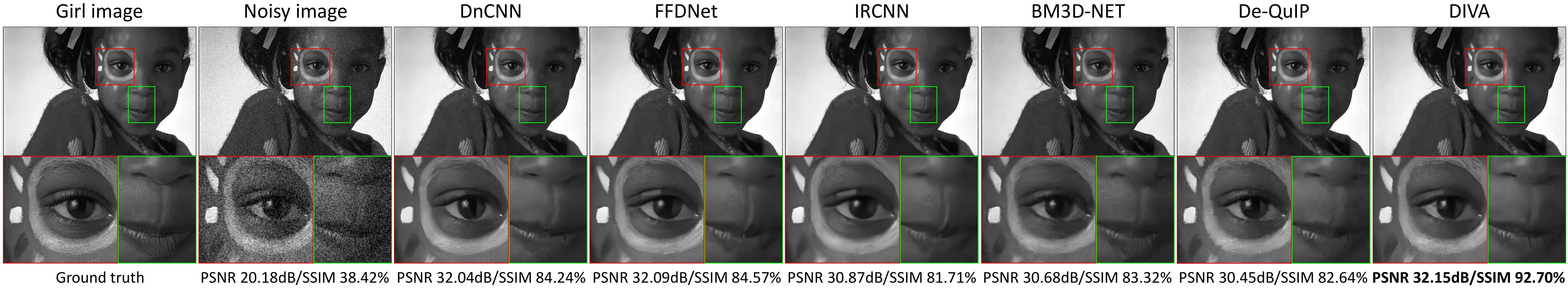}
\includegraphics[width=1\textwidth]{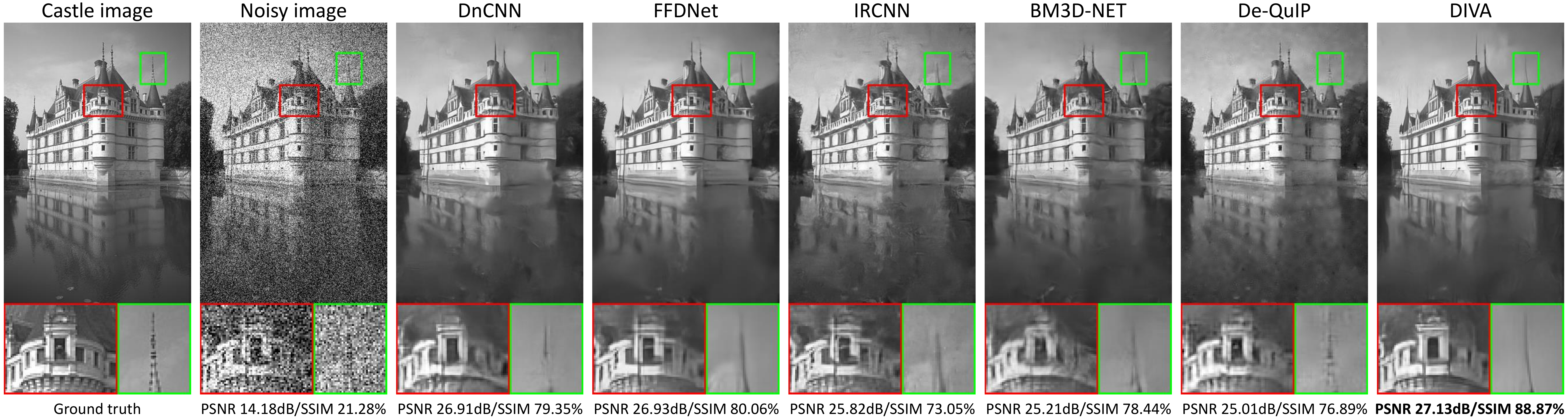}
\includegraphics[width=1\textwidth]{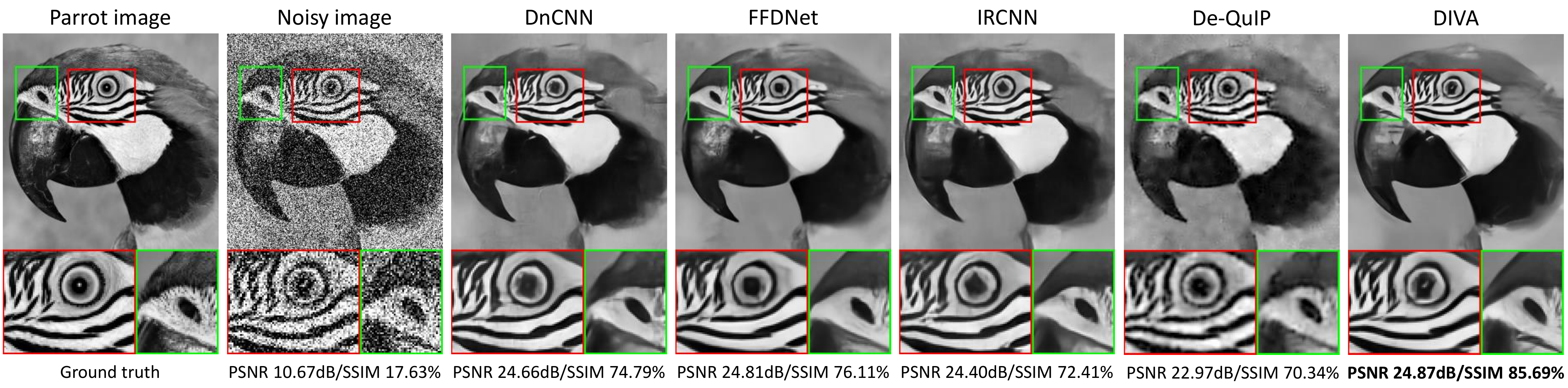}

\end{centering}
%%\vspace{-4mm}
\caption{Denoising image results using different methods. The \textit{Girl} image (Top), \textit{Castle} image (middle), and \textit{Parrot} image (bottom) are respectively contaminated with AWGN with $\sigma = 25$, $\sigma = 50$, and $\sigma = 75$.}
%\vspace{-2mm}
\label{fig:image_denoising}
\end{figure*}

Fig.~\ref{fig:image_denoising} illustrates denoising results for three images, \textit{Girl}, \textit{Castle} and \textit{Parrot}, from three datasets, for $\sigma =$ 25, 50 and 75 respectively. The qualitative analysis of the denoised images confirms the superiority of the proposed model.
\sd{Indeed, all competing methods fail to recover the original textures around the eye and lips in \textit{Girl} image, the sharp edges and peaks around the windows and roof in \textit{Castle} image, and the patterns in \textit{Parrot} image. IRCNN restores blurred edges, and BM3D-NET and De-QuIP generate some small artifacts. DnCNN and FFDNet give comparable PSNR, but low SSIM, caused by over-smoothed results, which were not able to retrieve small details. In contrast, \dk{DIVA} is faithful to the ground truths and restores the images with the right consistency by capturing the subtle details more reliably.}

%Indeed, for the \textit{Girl} image, all competing methods fail to recover the sclera, eyelashes and the patterns on the lips. IRCNN restores blurred edges, and BM3D-NET and De-QuIP generate some small artifacts around the eye and lips. DnCNN and FFDNet give comparable PSNR, but low SSIM, caused by over-smoothed results resulting into eye sclera, eyelashes, and lips lacking their original textures. In contrast, \dk{DIVA} is faithful to the ground truth and restores the girl's face with the right consistency. In \textit{Castle} image, IRCNN, BM3D-NET and De-QuIP fail to restore the sharp edges properly and introduce different artifacts. Among the comparison methods, DnCNN and FFDNet are the best, but were not able to retrieve small details, such as decorations around the windows and sharp peaks on the roof, wlile our model captures these details more reliably. Finally, in \textit{Parrot} image, IRCNN and De-QuIP generates several artifacts due to the presence of strong noise. 
%\dk{DIVA} preserves the patterns around the eye, nose, and parrot's beak more efficiently compared to DnCNN and FFDNet methods.

Visual and quantitative inspections indicate that \dk{DIVA} model conclusively outperforms its baseline method De-QuIP, as well as other advanced DL methods by a significant margin in terms of PSNR and SSIM. The DnCNN and FFDNet are the closest to \dk{DIVA}, but struggle to preserve image textures accurately, mainly because of a smoothing effect. \dk{DIVA} preserves most of the image fragments and textures in a better way without creating any visible artifacts and thus provides a denoised image closer to the ground truth.
%This justifies the conservation of the adaptive nature of the original De-QuIP scheme in our proposed unfolded model by embedding the quantum interaction and Hamiltonian architecture therein. This adaptability is even further enhanced by harnessing the power of convolutional layers.

% Moreover, notable improvement in SSIM values can be observed for low as well as for high-intensity noise.

\begin{table}[t!]
\setlength\tabcolsep{1.5pt}

\begin{scriptsize}

\begin{center}
\caption{Deblurring results in terms of average PSNR(dB) and SSIM(\%) values for \sd{two datasets degraded with three GB kernels and AWGN.}}
%\vspace{-2mm}
\label{tab:tab_psnr_ssim_gaussian_deblur}
\begin{tabular}{lcc cccc}
%%%%%%%%%%%%%%%%%%%%%%%%%Noise = 20%%%%%%%%%%%%%%%%%%%%%%%%%%%
\hline

Dataset & kernel$_{\sigma}$ & noise$_{\sigma}$ & \multicolumn{4}{c}{Methods}\\ 

& & & IDD-BM3D\cite{Danielyan2012bm3d} & Son \textit{et al.}\cite{Son2017fast} & DEBCNN\cite{Wang2018training} & \dk{DIVA}-A \\
\hline
%\cline{2-8}

%---------------------------------------------------

\multirow{3}{*}{BSD100}

& 1.6 & 2  & 27.17/86.14 & 23.18/73.47 & \bdb{28.47/87.90} & \bdr{29.97/89.65} \\

&  3  & 10 & 24.16/76.66 & 22.88/68.14 &\bdb{ 25.34/78.11} & \bdr{26.57/80.16} \\

&  5  & 10 & 22.75/71.74 & 22.17/65.92 & \bdb{22.79/71.94} & \bdr{23.73/74.09} \\

%\cline{2-8}
\hline

%---------------------------------------------------

\multirow{3}{*}{Set16}

& 1.6 & 2  & 30.85/93.41 & 29.87/93.29 & \bdb{31.34/94.39} & \bdr{32.38/95.37} \\

&  3  & 10 & 26.37/85.78 & 25.20/82.34 & \bdb{26.93/86.91} & \bdr{27.38/89.31} \\

&  5  & 10 & 24.23/82.24 & 23.63/80.55 & \bdb{27.28/82.76} & \bdr{28.11/87.22} \\

%\cline{2-8}
\hline

%---------------------------------------------------

\end{tabular}\end{center}
\end{scriptsize}
%\vspace{-5mm}
\end{table}

\begin{table}[t!]
\setlength\tabcolsep{2pt}

\begin{tiny}
\begin{center}
\caption{Deblurring results in terms of average PSNR(dB) and SSIM(\%) values for four datasets degraded with standard MB kernels and AWGN.
%The symbol -/- denotes that the results were not provided \sd{in the original paper} for a particular experiment.
}
%\vspace{-4mm}
\label{tab:tab_psnr_ssim_motion_deblur}
\begin{tabular}{l c cccccc}
%%%%%%%%%%%%%%%%%%%%%%%%%Noise = 20%%%%%%%%%%%%%%%%%%%%%%%%%%%
\hline

Dataset & noise$_{\sigma}$ & \multicolumn{6}{c}{Methods}\\

& & IDD-BM3D\cite{Danielyan2012bm3d} & FDN\cite{Kruse2017learning} & VEMNet\cite{Nan2020variational} & DWDN\cite{Dong2020deep} & DRED-DUN\cite{Kong2022deep} & \dk{DIVA}-A \\
\hline
%---------------------------------------------------

\multirow{4}{*}{Set10}

& 0    & 36.24/89.24  & -/- & -/- & \bdr{43.95}/\bdb{96.49} & \bdb{43.67}/96.38 & 43.54/\bdr{96.67} \\

& 2.55 & 30.75/86.63  & -/- & 31.71/89.95 & \bdr{33.28}/\bdb{93.12} & \bdb{33.16}/92.97 & 33.03/\bdr{93.54} \\

& 7.65 & 27.25/77.76  & -/- & 28.27/82.51 & \bdb{29.61}/88.07 & \bdr{29.80}/\bdb{88.48} & 29.38/\bdr{90.03} \\

& 12.75 & 25.71/71.38 & -/- & 26.62/77.68 & 26.92/83.16 & \bdr{27.49}/\bdb{84.05} & \bdb{27.42}/\bdr{85.79} \\

%\cline{2-8}
\hline

%---------------------------------------------------

\multirow{4}{*}{Levin}

& 0 & 37.48/94.68  & -/-  & -/- & \bdb{46.13/97.63} & 45.56/97.27 & \bdr{46.19/97.76} \\

& 2.55 & 33.75/92.19  & 34.05/93.35 & 34.31/94.31 & \bdr{36.90/96.14} & 36.02/95.79 & \bdb{36.19/95.86} \\

& 7.65 & 29.26/85.78  & 29.77/85.83 & 30.50/87.86 & 32.77/91.79 & \bdb{32.87/91.97} & \bdr{33.12/92.46} \\

& 12.75 & 27.33/78.92 & 27.94/81.39 & 28.52/82.73 & 30.77/88.57 & \bdr{30.89}/\bdb{88.79} & \bdb{30.80}/\bdr{89.87} \\

%\cline{2-8}
\hline

%---------------------------------------------------

\multirow{4}{*}{Sun \textit{et al.}}

& 0    &  37.14/90.42  & -/- & -/- & \bdr{43.10}/\bdb{97.19}     & 42.49/97.08 &  \bdb{42.65}/\bdr{97.36} \\

& 2.55 & 32.24/87.79  & 32.63/88.87 & 32.73/90.13 & 34.05/92.25 & \bdb{34.43/92.97} & \bdr{34.44/93.49} \\

& 7.65 & 28.74/77.86  & 28.97/78.42 & 29.41/81.08 & 29.11/86.31     & \bdb{29.88/87.28} & \bdr{30.30/89.14} \\

& 12.75 & 27.30/73.24 & 27.62/74.52 & 28.04/77.89 & 27.81/80.85     & \bdr{28.20}/\bdb{81.59} & \bdb{27.95}/\bdr{83.36} \\

%\cline{2-8}
\hline

%---------------------------------------------------

\multirow{4}{*}{Set12}

& 0     & -/-  & -/-  & -/-  & -/-  & -/-  &  \bdr{43.48/96.39} \\

& 2.55  & 31.43/88.14 & 31.43/89.17 & \bdb{31.93/90.19}  & -/-  & -/-  & \bdr{33.77/92.58}  \\

& 7.65  & 27.56/80.09 & 27.89/80.86 & \bdb{28.47/82.78} & -/-  & -/-  & \bdr{28.97/87.89} \\

& 12.75 & 25.95.74.88 & 26.28/76.24 & \bdb{26.77/78.13} & -/-  & -/-  & \bdr{27.28/84.45} \\

%\cline{2-8}
\hline

%---------------------------------------------------

\end{tabular}\end{center}
\end{tiny}
\textit{$\ast$The symbol -/- denotes that the results were not provided in the original paper for a particular experiment.}
%\vspace{-5mm}
\end{table}

%\vspace{-2mm}
\subsubsection{Image Deblurring}
\label{sec:result_deblur}
%\vspace{-1mm}

Image deblurring results for GB are illustrated on two benchmark datasets degraded with three different GB kernel settings of size $25 \times 25$: (i) GB kernel with standard deviation of 1.6 and AWGN with $\sigma=2$,
(ii) GB kernel with standard deviation of 3 and AWGN with $\sigma=10$,
(iii) GB kernel with standard deviation of 5 and AWGN with $\sigma=10$.
Table~\ref{tab:tab_psnr_ssim_gaussian_deblur} regroups the average PSNRs and SSIMs obtained by all competing methods. One can observe that the benchmark DEBCNN \cite{Wang2018training} method performs much better than the model-based IDD-BM3D \cite{Danielyan2012bm3d} and learning-based Son \textit{et al.} \cite{Son2017fast} schemes. \dk{DIVA}-A outperforms DEBCNN by 1.1 dB in PSNR and 2\% in SSIM and 0.8 dB in PSNR and 2.6\% in SSIM on average for BSD100 and Set16 datasets, respectively.

\sd{In Fig.~\ref{fig:image_deblurG}, a qualitative evaluation shows that the proposed method not only generates better image contrast but also retrieves sharp edges with more details than the other approaches, like IDD-BM3D and Son \textit{et al.} \cite{Son2017fast}, where random artifacts and blurred edges are visible in the deblurred outputs.
Our DL model restores the \textit{Penguin} image with much sharper and more precise edges than the DEBCNN, for which edges look hazy.
Thus, though DEBCNN and \dk{DIVA}-A are the two best models in this setting, our model uniformly outperforms the sophisticated DEBCNN method for GB problems.}

% add figures  --------------------------------------
\begin{figure}[t!]
\begin{centering}

\includegraphics[width=0.5\textwidth]{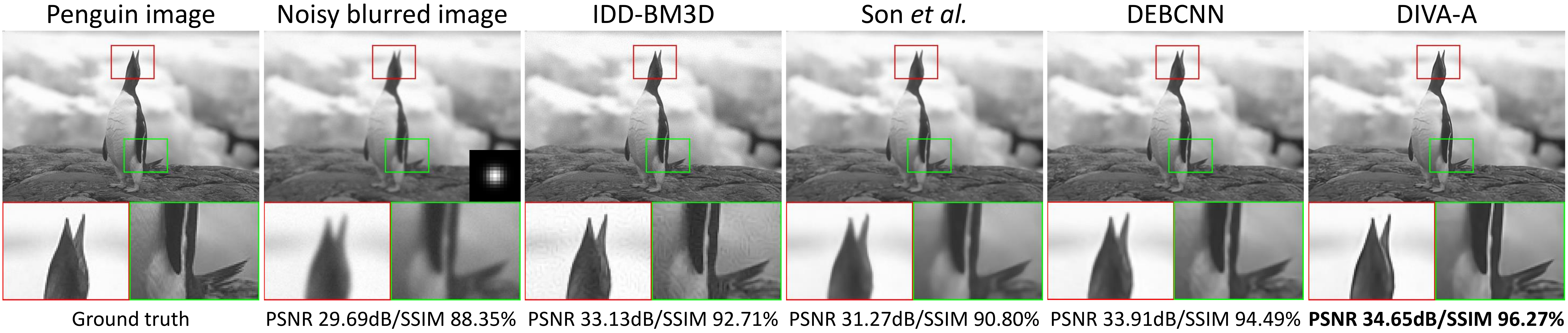}

\end{centering}
%\vspace{-3mm}
\caption{Image deblurring results for \textit{Penguin} image degraded by a $25 \times 25$ GB kernel of standard deviation 1.6 with random AWGN of standard deviation $2$.}
%\vspace{-5mm}
\label{fig:image_deblurG}
\end{figure}

%Two deblurred images using the different methods are presented in Fig.~\ref{fig:image_deblurG}. A qualitative evaluation shows that the proposed method not only generates better image contrast but also retrieves sharp edges with more details than the other approaches. For example, in \textit{Horse} and \textit{Penguin} images, IDD-BM3D produces better contrast, but random patterns are visible in the deblurred outputs, whereas Son \textit{et al.} \cite{Son2017fast} fails to preserve sharp edges and the overall images appear blurred compared to others. In the \textit{Penguin} image, our DL model restores the beak, tail, and flipper of the penguin with much sharper and more precise edges than the DEBCNN, for which edges look hazy. A similar observation can be drawn from the \textit{Horse} image, where the overall head looks much sharper by our method. Thus, though DEBCNN and \dk{DIVA}-A are the two best models in this setting, our model uniformly outperforms the sophisticated DEBCNN method for Gaussian deblurring problems.

Table~\ref{tab:tab_psnr_ssim_motion_deblur} gives the average deblurring performance of our method in terms of PSNRs and SSIMs in contrast to other standard models from the literature under eight commonly used MB kernels \cite{Levin2011efficient, Kong2022deep} and four different noise levels. One should note that the code or trained models provided by the authors are used to generate these results. As the first observation, one can see that DWDN and DRED-DUN outperform the conventional IDD-BM3D, FDN and VEMNet for the Set10, Levin and Sun \textit{et al.} datasets, which is consistent with the findings in \cite{Kong2022deep}. \sd{Secondly, DWDN performs} better in the case of low/no noise in terms of PSNRs compared to DRED-DUN and our proposed model. DRED-DUN is more accurate for high levels of noise. On the contrary, our proposed model exhibits the best SSIMs with a gain up to 0.15-1.8\% against the DWDN and DRED-DUN for low as well as high noise levels and this efficiency increases with noise intensity. In terms of PSNR values, our model often stays in the top two and only fails to do so for Set10, where the average PSNR gaps between the \sd{best two methods and our model is very small}. Noticeably, although \dk{DIVA}-A sometimes offers slightly worse PSNRs than DWDN and DRED-DUN, it requires only \sd{half of the tunable parameters (shown in Fig.~\ref{subfig:Time_Parameter_deblur}).}
% and significantly reduces the computation and running costs, as shown in Fig.~\ref{subfig:Time_Parameter_deblur}.
Finally for Set12, our model unilaterally dominates the comparison and exceeds its nearest rival VEMNet by up to an average of 1dB PSNR and 3.5\% SSIM. %Hence,  \dk{DIVA}-A is one of the top two-performing models for motion deblurring problems most of the time.

% add figures  --------------------------------------
\begin{figure}[t!]
\begin{centering}

\includegraphics[width=0.5\textwidth]{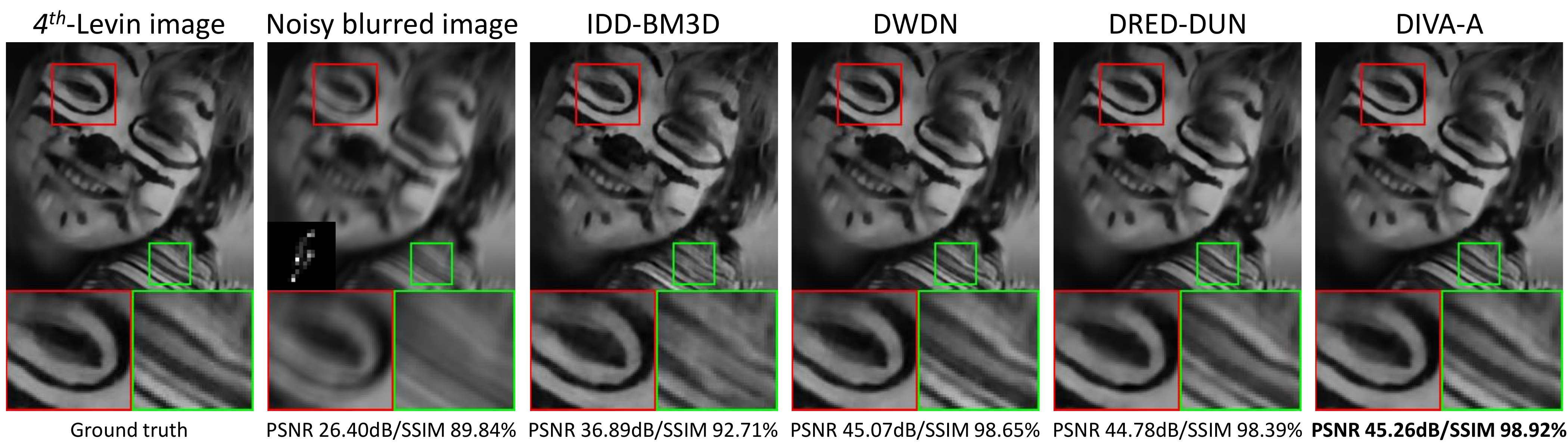}

\includegraphics[width=0.5\textwidth]{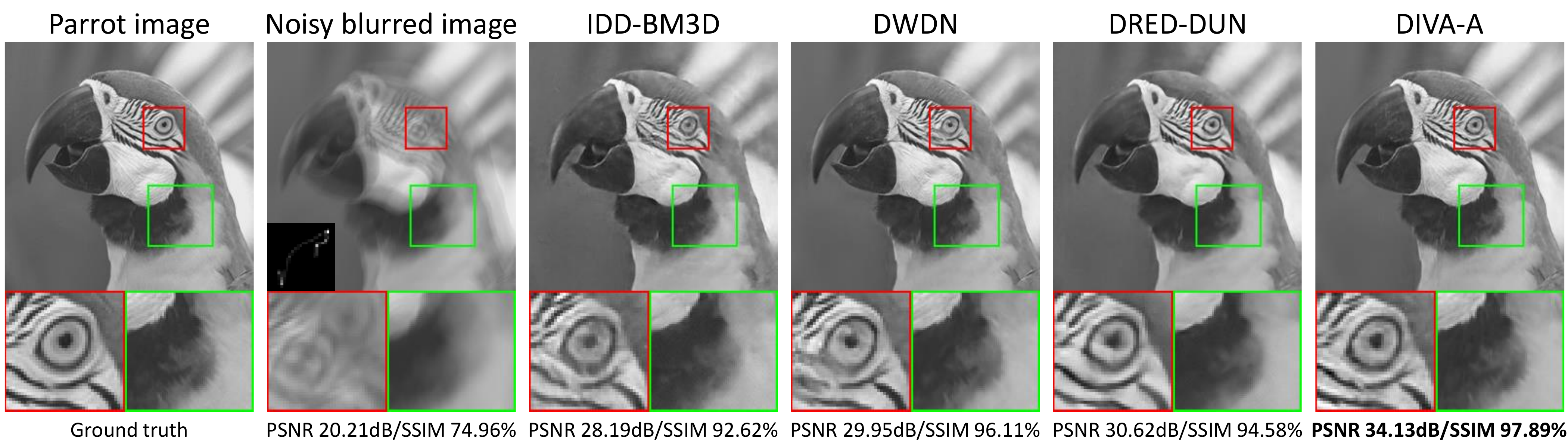}

\end{centering}
%\vspace{-3mm}
\caption{Deblurring results for MB kernels.
The first row shows restored \textit{4th}-image from the Levin dataset with $17 \times 17$ MB kernel, and second row shows restored \textit{Parrot} images with $25 \times 25$ MB kernel.}
%\vspace{-5mm}
\label{fig:image_deblurM1}
\end{figure}

% add figures  --------------------------------------
\begin{figure*}[t!]
\begin{centering}

\includegraphics[width=1\textwidth]{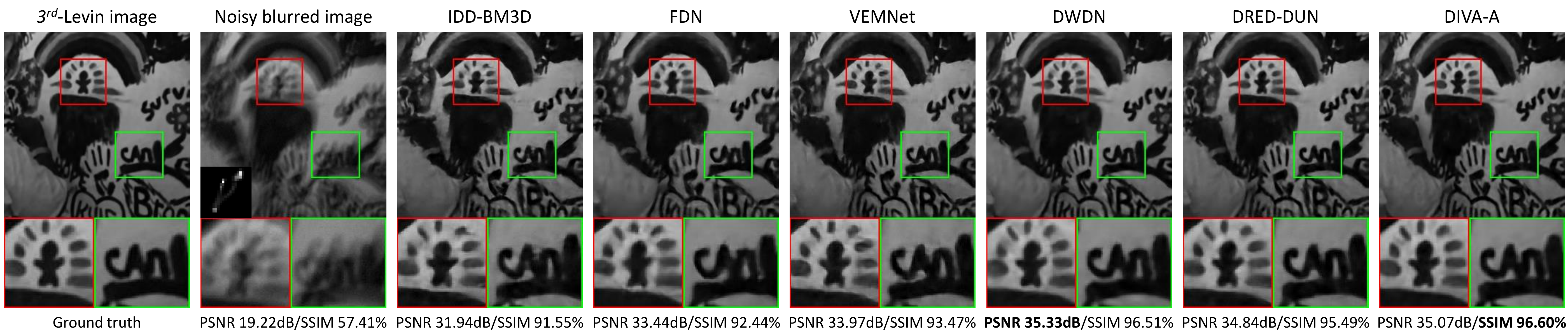}

\end{centering}
%\vspace{-2mm}
\caption{Deblurring results for the \textit{3rd}-image from the Levin dataset with motion blur kernel of size $23 \times 23$ and random AWGN $\sigma=2.55$.}
%\vspace{-5mm}
\label{fig:image_deblurM2}
\end{figure*}

\sd{For visual assessment, restored images from images degraded with three different MB kernels are shown in Figs.~\ref{fig:image_deblurM1},~\ref{fig:image_deblurM2}. Fig.~\ref{fig:image_deblurM1} shows that for the \textit{4th}-Levin and \textit{Parrot} images respectively under a MB kernel of size $17 \times 17$ and $25 \times 25$, the quality of the restored images by our model is considerably improved compared to the other methods.
In particular, the finer texture of the images is severely smoothed out by IDD-BM3D, DWDN and DRED-DUN, as shown in the zoomed boxes. 
Furthermore, the overall visual impression of the restored images is improved, as visible on the facial decorations and minute patterns that are better preserved with \dk{DIVA}-A. Finally, Fig.~\ref{fig:image_deblurM2} offers a similar conclusion for the restored \textit{3th}-Levin image under a $23 \times 23$ MB kernel with AWGN with $\sigma=2.55$.
%Finally, Fig.~\ref{fig:image_deblurM2} regroups the restored \textit{3th}-Levin image under a $23 \times 23$ MB kernel with AWGN with $\sigma=2.55$. Our model is better equipped to retrieve these small precise details like decorations, edges and texts than the other methods.
Thus, under MB kernels \dk{DIVA}-A demonstrates a better efficiency in recovering edges and patterns of the original images. DWDM and DRED-DUN produce comparative results compared to our model, but with lower contrast. Quantitatively, our method is always among the best two approaches in this context.}

\begin{table}[t!]
\setlength\tabcolsep{1.7pt}

\begin{scriptsize}
\begin{center}
\caption{SR results in terms of average PSNR(dB) and SSIM(\%) values for 4 benchmark datasets degraded with bicubic downsampling with downsampling factors of 2, 3 and 4.}
%\vspace{-2.5mm}
\label{tab:tab_psnr_ssim_BiCuSupReso}
\begin{tabular}{l c ccccc}
\hline

Dataset & Scale & \multicolumn{5}{c}{Methods}\\ 

& & LapSRN\cite{Lai2017deep} & MemNet\cite{Tai2017MemNet} & CARN\cite{Ahn2018fast} & DRLN\cite{Anwar2022densely} & \dk{DIVA}-A \\
\hline

%---------------------------------------------------

\multirow{3}{*}{Set5}

& 2x & 37.52/95.91 & \bdb{37.78}/95.97 & 37.76/95.90 & \bdr{38.27}/\bdb{96.16} & 37.42/\bdr{97.43} \\

& 3x & 33.82/92.27 & 34.09/92.48 & \bdb{34.29}/92.55 & \bdr{34.78}/\bdb{93.03} & 33.14/\bdr{93.36} \\

& 4x & 31.54/88.50 & 31.74/88.93 & \bdb{32.13/89.37} & \bdr{32.63/90.02} & 30.87/\bdr{90.02} \\

\hline

%---------------------------------------------------

\multirow{3}{*}{Set14}

& 2x & 33.08/91.30 & 33.28/91.42 & 33.52/91.66 & \bdr{34.28}/\bdb{92.31} & \bdb{33.67}/\bdr{93.69} \\

& 3x & 29.87/83.20 & 30.00/83.50 & \bdb{30.29}/84.07 & \bdr{30.73}/\bdb{84.88} & 29.18/\bdr{85.34} \\

& 4x & 28.19/77.20 & 28.26/77.23 & \bdb{28.60}/78.06 & \bdr{28.94}/\bdb{79.00} & 27.74/\bdr{80.66} \\

\hline

%---------------------------------------------------

\multirow{3}{*}{BSD100}

& 2x & 31.80/89.50 & 32.08/89.78 & \bdb{32.09}/89.78 & \bdr{32.44}/\bdb{90.28} & 32.00/\bdr{90.49} \\

& 3x & 28.82/79.80 & 28.96/80.01 & \bdb{29.06}/80.34 & \bdr{29.36}/\bdb{81.17} & 28.91/\bdr{82.15} \\

& 4x & 27.32/72.80 & 27.40/72.81 & 27.58/73.49 & \bdr{27.83}/\bdb{74.44} & \bdb{27.66}/\bdr{76.95} \\

%\cline{2-8}
\hline

%---------------------------------------------------

\multirow{3}{*}{Urban100}

& 2x & 30.41/91.00 & 31.31/91.95 & \bdb{31.51/93.12} & \bdr{33.37/93.90} & 31.48/93.06 \\

& 3x & 27.07/82.80 & \bdb{27.56}/83.76 & 27.38/84.04 & \bdr{29.21/87.22} & 27.54/\bdb{85.31} \\

& 4x & 25.21/75.60 & 25.50/76.30 & \bdb{26.07}/78.37 & \bdr{26.98}/\bdb{81.19} & 25.39/\bdr{81.29} \\

%\cline{2-8}
\hline

%---------------------------------------------------
\end{tabular}\end{center}
\end{scriptsize}
%\vspace{-5mm}
\end{table}

%\vspace{-2mm}
\subsubsection{Single Image Super-Resolution (SR)}
\label{sec:result_sr}

This subsection presents SR results for two standard downsampling operators, bicubic downsampling (BD) and Gaussian downsampling (GD). Tables~\ref{tab:tab_psnr_ssim_BiCuSupReso},~\ref{tab:tab_psnr_ssim_GausSupReso} regroup average PSNR and SSIM values of different methods on four datasets for BD and GD respectively. One may observe that the recently introduced benchmark method DRLN \cite{Anwar2022densely} provides the best performance in both contexts. DRLN has a complex network architecture with dense residual Laplacian modules powered by 34 million parameters. In contrast, the proposed model has a much simpler architecture, and requires only 850K parameters approximately. Nevertheless, our model obtains the best SSIM for three datasets (\textit{e.g.}, Set5, Set14 and BSD100) and among the top two SSIM for Urban100 images for BD. One can see an average gain of 1.5\% SSIM by our method over DRLN in the BD scenario. For GD problems, our method struggles to produce competitive results against benchmark DRLN, RDN and DFAN approaches. Note that for SR problem our method upsamples the observed LR image by bicubic interpolation to obtain an initial HR image before enhancing it using the trained DL network.

The visual inspections of Figs.~\ref{fig:image_super_resoBx3},~\ref{fig:image_super_resoBx4} and \ref{fig:image_super_resoGx2} illustrate the potential of our method for SR. Figs.~\ref{fig:image_super_resoBx3} and \ref{fig:image_super_resoBx4} correspondingly display the restored HR images from the LR BD \textit{Zebra} and \textit{Baby-face} images with scale factors of 3 and 4. The visual effects of HR images recovered by our method are better than others and higher in accuracy.
\sd{For example, in our retrieved HR images the stripes on the zebra's body, in \textit{Baby-face} image the textures and shapes of eye, lips and nose have better specifications than the other methods.
Fig.~\ref{fig:image_super_resoGx2} shows the reconstructed HR images from the LR \textit{Book-cover} image obtained by GD with scale factor of 2. Observation reveals that our method efficiently recovers the edges, patterns and texts of the original image from LR data. Moreover, our method strongly competes with the benchmark DRLN and beats it in some respects, especially in terms of overall visual quality and preservation of the image structure.}

%The visual inspections of Figs.~\ref{fig:image_super_resoBx3},~\ref{fig:image_super_resoBx4} and \ref{fig:image_super_resoGx2} illustrate the potential of our method for SR. Figs.~\ref{fig:image_super_resoBx3} and \ref{fig:image_super_resoBx4} correspondingly display the restored HR images from the LR BD \textit{Zebra} and \textit{Baby-face} images with scale factors of 3 and 4. The visual effects of HR images recovered by our method are better than others and higher in accuracy. For example, in our retrieved HR \textit{Zebra} image the stripes on the zebra's body, in \textit{Baby-face} image the textures of eyelashes and eyeball, and lips and nose shapes have better specifications than the other methods. Fig.~\ref{fig:image_super_resoGx2} shows the reconstructed HR images from the LR \textit{Book-cover} image obtained by GD with scale factor of 2. Observation reveals that our method efficiently recovers the edges, patterns and texts of the original image from LR data. Moreover, our method strongly competes with the benchmark DRLN and beats it in some respects, especially in terms of overall visual quality and preservation of the image structure. Here again, the interaction layer equipped our network in an efficient way to gather image features/structures despite having a simplified network backbone. Thus, harnessing this non-local blueprint, our DL network recovers the HR images with better visual salient attributes like small scale patterns, sharp edges and textures for super-resolution problems.

\begin{table}[t!]
\setlength\tabcolsep{1.7pt}

\begin{scriptsize}
\begin{center}
\caption{SR results in terms of average PSNR(dB) and SSIM(\%) values for 4 benchmark datasets degraded with GD by using a $7 \times 7$ GB kernel of standard deviation $1.6$ with scaling factors of 2, 3 and 4.}
%\vspace{-4mm}
\label{tab:tab_psnr_ssim_GausSupReso}
\begin{tabular}{l c ccccc}
\hline

Dataset & Scale & \multicolumn{5}{c}{Methods}\\ 

& & IRCNN\cite{Zhang2017learning} & DFAN\cite{Li2022DFAN} & RDN\cite{Zhang2021residual} & DRLN\cite{Anwar2022densely} & \dk{DIVA}-A \\
\hline

%---------------------------------------------------

\multirow{3}{*}{Set5}

& 2x & \bdr{35.34}/\bdb{93.04} & -/- & -/- & -/- & \bdb{33.62}/\bdr{93.79} \\

& 3x & 33.38/91.82 & 34.50/92.74 & \bdb{34.58/92.80} & \bdr{34.81/92.97} & 32.70/91.45\\

& 4x &  \bdr{30.76}/\bdb{85.47} & -/- & -/- & -/- & \bdb{29.02}/\bdr{85.76}\\

\hline

%---------------------------------------------------

\multirow{3}{*}{Set14}

& 2x & \bdr{31.98}/\bdb{88.49} & -/- & -/- & -/- & \bdb{30.88}/\bdr{90.65} \\

& 3x & 29.63/82.81 & 30.43/84.19 & \bdb{30.53/84.47} & \bdr{30.81/84.87} & 28.97/83.47\\

& 4x & \bdr{27.73}/\bdb{74.12} & -/- & -/- & -/- & \bdb{26.86}/\bdr{76.01} \\

\hline

%---------------------------------------------------

\multirow{1}{*}{BSD100}

& 3x & 28.65/79.22 & 29.17/80.58 & \bdb{29.23/80.79} & \bdr{29.40/81.21} & 28.26/80.65\\

\hline

%---------------------------------------------------

\multirow{1}{*}{Urban100}

& 3x & 26.77/81.54 & 28.27/85.26 & \bdb{28.46/85.82} & \bdr{29.11/86.97} & 27.72/84.92\\

\hline

%---------------------------------------------------
\end{tabular}\end{center}
\end{scriptsize}
%\vspace{-5mm}
\end{table}

% add figures  --------------------------------------
\begin{figure*}[t!]
\begin{centering}

\includegraphics[width=1\textwidth]{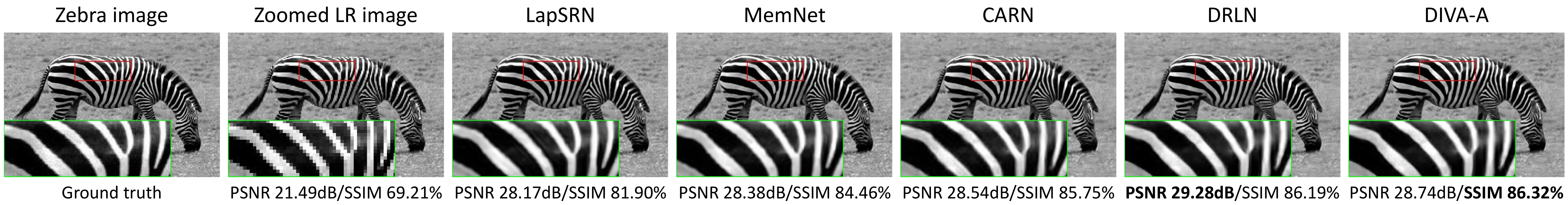}

\end{centering}
%\vspace{-2mm}
\caption{SR results for \textit{Zebra} image for a bicubic downsampling with scaling factor 3.}
\label{fig:image_super_resoBx3}
%\vspace{-4mm}
\end{figure*}

\begin{figure*}[t!]
\begin{centering}

\includegraphics[width=1\textwidth]{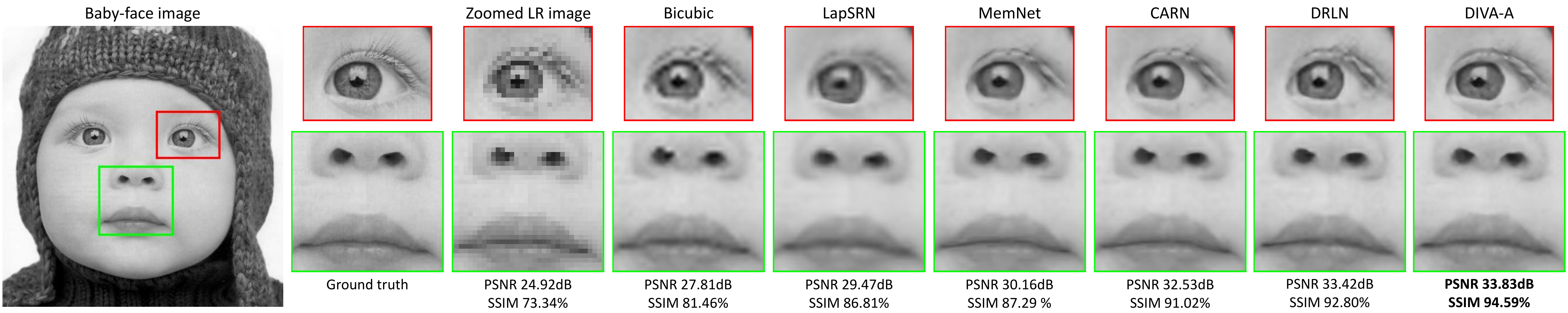}

\end{centering}
%\vspace{-2mm}
\caption{Two zoomed regions of the restored HR \textit{Baby-face} images, extracted from SR results for a bicubic downsampling with scaling factor 4.}
\label{fig:image_super_resoBx4}
%\vspace{-4mm}
\end{figure*}

% add figures  --------------------------------------
\begin{figure}[t!]
\begin{centering}

\includegraphics[width=0.5\textwidth]{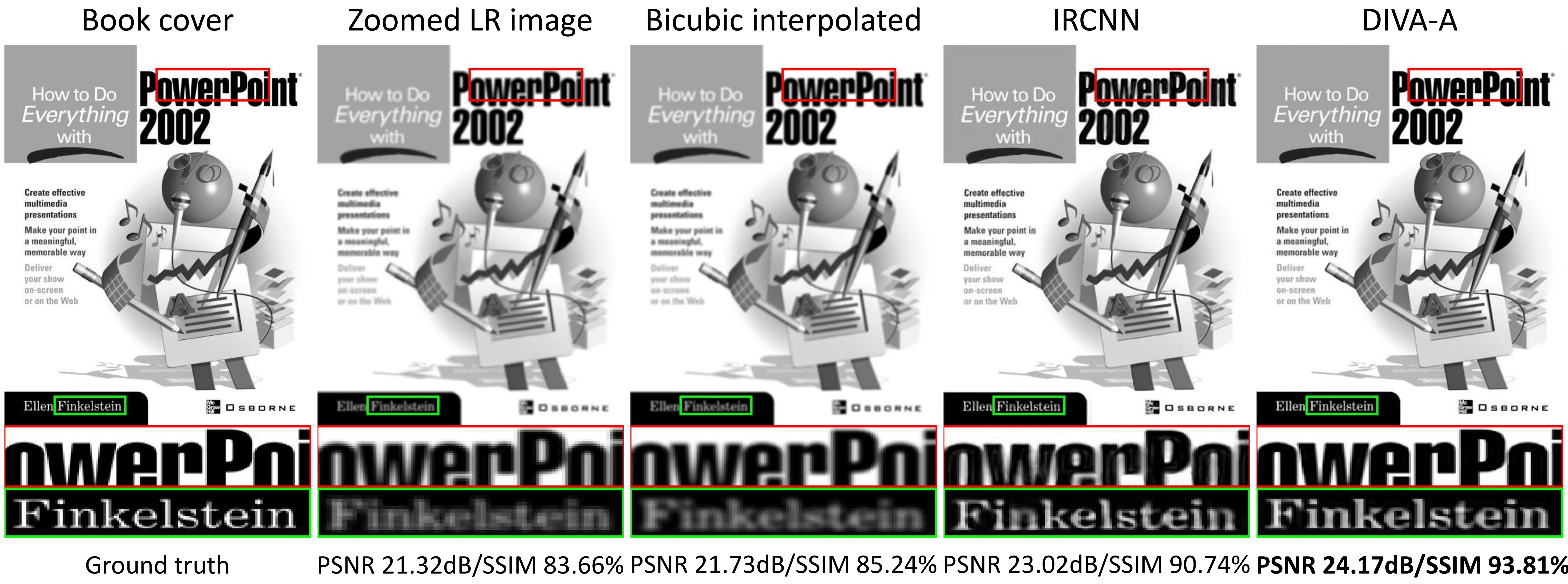}

\end{centering}
%\vspace{-2mm}
\caption{Restored HR \textit{Book-cover} images from LR images generated by GD under a $7 \times 7$ GB kernel of standard deviation 1.6 with scaling factor 2.}
\label{fig:image_super_resoGx2}
%\vspace{-4mm}
\end{figure}

%\vspace{-2mm}
\subsubsection{Image Inpainting}
\label{sec:result_inpaint}

Table~\ref{tab:tab_psnr_ssim_image_impainting} illustrates our model performance in terms of average PSNR and SSIM on Set5 and Set12 datasets compared to the standard IRCNN network for image inpainting problems. Our model outperforms IRCNN in almost all situations when 20\%, 50\%, and 80\% of random pixels are missing in the degraded observed images. \sd{DIVA-A provides a improvement of 0.2-1 dB in PSNR and 0.6-5.5\% in SSIM over IRCNN and this gain increases with data lacking.}

%For a small amount of missing pixels, \dk{DIVA}-A model provides a slight improvement of 0.2 dB in PSNR and 0.6\% in SSIM over IRCNN, while a gain of 1 dB in PSNR and 5.5\% in SSIM was obtained for 80\% of data lacking.

The visual analysis of Fig.~\ref{fig:image_inpainting} confirms the quantitative results. \sd{From the restored \textit{F-16 Jet} image, it appears that our model efficiently reproduces the F-16 logo, borders and sharp edges \sd{despite 50\% of data missing}, whereas IRCNN fails to do so and loses/distorts many details in the restored output. 
Hence, our model can gather local information from the image neighborhood quite promisingly and delivers a high-quality restored image even with limited pixels available.}
\footnote{More visual results can be found in the supplementary material.}

%The visual analysis of Fig.~\ref{fig:image_inpainting} confirms the quantitative results. From the restored \textit{F-16 Jet} image, it appears that our model efficiently reproduces the F-16 logo on the jet's tail, whereas IRCNN fails to do so. Similarly, in the \textit{Boat} image, despite 80\% of data missing our model recovers minute details like the ropes and structures on the deck. On the contrary, the image restored by IRCNN \dk{is} more blurry and loses/distorts many details, such as the borders, sharp edges, and ropes in the restored output. Hence, our model can gather local information from the image neighborhood quite promisingly through the quantum interaction layer and delivers a high-quality restored image even with limited pixels available.

%\vspace{-4mm}
\section{Discussions}
\label{sec:discuss}

In this section, we briefly recap the benefits and limitations of our proposed networks and future prospects in this regard.

\textit{Advantages:}
With the quantum principles of the baseline De-QuIP algorithm, our proposed \dk{DIVA}/\dk{DIVA}-A network provides an efficient DL method for image restoration following the deep unfolding philosophy. Indeed, the \sd{use} of quantum concepts like patch interaction layer and Hamiltonian kernel \sd{makes} our models better equipped than others. The local structure/similarities in an image neighborhood are preserved through the interaction layer exploiting the local patch groups that convey an intrinsic non-local network architecture. \dk{Processing of the local information by this interaction layer significantly enhances the performances of the network. It even yields  a smaller network depth, leading to a good trade-off between the performance and computational cost, as portrayed in Sec.~\ref{sec:ablation}. Harnessing the power of back-propagation, our networks uniquely tune all hyperparameters, such as proportionality constant, Planck constant and thresholding energy, for each patch. This enables network adaptability with several image restoration tasks, and leads to promising performances.}

\begin{table}[t!]

\begin{scriptsize}
\begin{center}
\caption{Image inpainting results in terms of average PSNR(dB) and SSIM(\%) values for two benchmark datasets for respectively 20\%, 50\% and 80\% pixels missing.}
%\vspace{-2mm}
\label{tab:tab_psnr_ssim_image_impainting}
\begin{tabular}{l c ccc}
%%%%%%%%%%%%%%%%%%%%%%%%%Noise = 20%%%%%%%%%%%%%%%%%%%%%%%%%%%
\hline
Dataset & Missing pixels' & Input & \multicolumn{2}{c}{Methods}\\ 

& & & IRCNN\cite{Zhang2017learning} & \dk{DIVA}-A \\
\hline
%\cline{2-8}

%---------------------------------------------------

\multirow{3}{*}{Set5}

& 20\% & 13.33/38.61  & 41.62/98.67 & \bdr{41.85/99.24} \\

& 50\% & 9.34/23.44  & 35.57/95.87 & \bdr{36.08/97.84} \\

& 80\% & 7.29/12.40  & 29.41/88.54 & \bdr{30.38/94.01} \\

%\cline{2-8}
\hline

%---------------------------------------------------

\multirow{3}{*}{Set12}

& 20\% & 12.46/27.93  & \bdr{39.06}/98.29 & 38.57/\bdr{99.15} \\

& 50\% & 8.48/14.45  & 32.82/94.53 & \bdr{33.02/97.21} \\

& 80\% & 6.44/6.71  & 26.75/84.53 & \bdr{27.73/91.92} \\

%\cline{2-8}
\hline

%---------------------------------------------------

\end{tabular}\end{center}
\end{scriptsize}
%\vspace{-6mm}
\end{table}

\textit{Limitations:}
In the case of a challenging image degradation task, our method may \dk{sometimes} struggle to produce a better recovered image than other benchmarks. To restore a Gaussian downsampled LR image, we notice that our DL model fails to compete in quantitative data against benchmark methods, like DRLN, RDN, and DFAN, as noted in Table~\ref{tab:tab_psnr_ssim_GausSupReso}. However, the overall visual efficiency of our method is quite good, as depicted in Fig.~\ref{fig:image_super_resoGx2}. Perhaps in presence of a strong decay, such as 'blur$+$downsampling', our method does not match the true pixels' intensity, which seems to be the main reason for the lower quantitative measures. Instead, our method utilizes the interaction layer to provide better visual quality by preserving the image structure, patterns, and textures with more \dk{details}. Furthermore, our proposed models are trained in an end-to-end supervised manner, \textit{i.e.}, we need the clean-degraded image pairs for training. %Thus, in absence of ground truth it becomes extremely difficult to create a training dataset.
\dk{However it is worth-noting that the proposed  method is much  simpler and not specialized in a specific task as  is the case for the other methods.}

\textit{Future perspectives:}
The quantum mechanics-based imaging methods open up a broad spectrum of future prospects. Following the limitations, the obvious direction \sd{would be} an unsupervised DL network design, that essentially solves the training data problem and extends our reach to real-life applications more \sd{reliably} \cite{Morsier2016kernel, Pereyra2017fast}.
% Chen2022Robust,
Another possibility is to design a versatile network by stacking the proposed \dk{DIVA} to build a deep and more complex architecture like UNet \cite{Kong2022deep} and offer some attention mechanisms \cite{Anwar2022densely} to make the patch interaction robust while preserving the core philosophy. This complex network system should enhance the capacity of the proposed network in challenging degradation operators and even for blind imaging problems. Furthermore, the idea of quantum interaction can also be treated as a transformer in a deep architecture \cite{Wang2022Uformer}. Another interesting prospect would be to explore imaging problems beyond the Gaussian model since baseline De-QuIP is well-adapted for such tasks without modifying the global architecture. Combining graph signal processing model with the proposed quantum-based interaction framework is also an interesting perspective \cite{Hua2019learning}.

% Another possibility is to design a versatile network by stacking the proposed \dk{DIVA} to build a deep and more complex architecture like UNet, and preserving the core philosophy of interaction offers some attention mechanisms.

% add figures  --------------------------------------
\begin{figure}[t!]
\begin{centering}

\includegraphics[width=.49\textwidth]{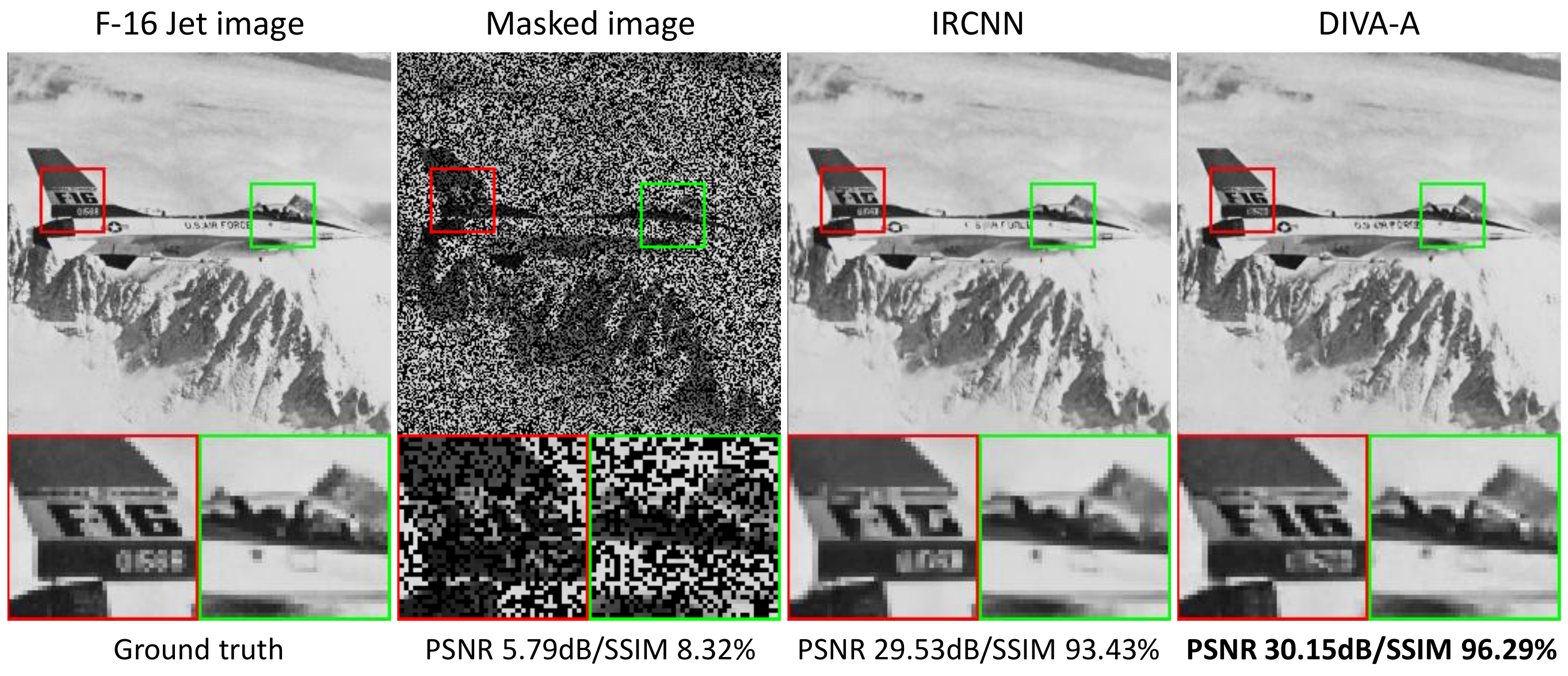}

\end{centering}
%\vspace{-2mm}
\caption{Restored \textit{F-16 Jet} images, when 50\% pixels' are missing.}
\label{fig:image_inpainting}
%\vspace{-5mm}
\end{figure}

% Image inpainting results by \dk{DIVA}-A. The first row shows restored \textit{F-16 Jet} images when 50\% pixels' are missing and the second row shows restored \textit{Boat} images when 80\% pixels' are missing.

%\vspace{-4mm}
\section{Conclusions}
\label{sec:conclusion}

This paper introduces a novel neural network approach to solve image denoising problems, further extended to general image restoration tasks relying on the philosophy of quantum many-body theory. Our model recasts the baseline De-QuIP algorithm into a DL framework and optimizes the relevant parameters by exploiting the power of back-propagation approach. The proposed unfolded CNN architecture inherently employs various quantum mechanical components, such as interaction and Hamiltonian operator, from its baseline method to boost up the network performance while significantly reducing the training cost.
%In a local image neighborhood, the quantum interactions give the patch similarity measures, whereas the local frequencies of the adaptive basis rely on the Hamiltonian operator.
Integration of these key features from the quantum theory enables our proposed model to be well-adapted for handling several imaging problems efficiently. We conduct thorough ablation investigations and present extensive assessments regarding the network design. Finally, we perform comprehensive evaluations of our proposed DL methods for various imaging problems, such as denoising, deblurring, single image super-resolution, and inpainting. In all cases, notable improvements were shown in the image restoration performance, especially overall visual quality, compared to standard well-established techniques from the literature.

%\vspace{-5mm}
\section*{Acknowledgment}
\label{sec:acknowledge}

Sayantan Dutta would like to thank Mr. Nishchal Prasad, Ph.D. student at Institut de Recherche en Informatique de Toulouse (IRIT), UMR CNRS 5505, Universit\'e de Toulouse, France, for his valuable inputs in this work.
Authors also thank CNRS for funding through the 80 prime program.
%\vspace{-4mm}

%\section*{References}

%\bibliographystyle{elsarticle-num}
\bibliographystyle{IEEEbib}
\bibliography{DIVA}

%%%%%%%%%%%%%%%%%%%%%%%%%%%%%%%%%%%%%%%%%%%%%%%%%%%%%%%%%%%%%%%%%%%%%%%%%%%% Supplementary Material: %%%%%%%%%%%%%%%%%%%%%%%%%%%%%%%%%%%%%%%%%%%%%%%%%%%%%%%%%%%%%%%%%%%%%%%%%%%%%%%%%%%%%%%

\newpage ~~ \newpage

\begin{onecolumn}

\begin{center}

\section*{\Huge{SUPPLEMENTARY MATERIAL:}\\
\Large{FOR}\\
\LARGE{DIVA: Deep Unfolded Network from Quantum Interactive Patches for Image Restoration}
}

\vspace*{5mm}

{Sayantan~Dutta$^{1,2,\ast}$,~Adrian~Basarab$^{3}$,~Bertrand Georgeot$^{2}$,~and~Denis~Kouam\'e$^{1}$}

\vspace{.2cm}
$^{1}$Institut de Recherche en Informatique de Toulouse, UMR CNRS 5505, Universit\'e de Toulouse, France

\vspace{.1cm}
$^{2}$Laboratoire de Physique Th\'eorique, Universit\'e de Toulouse, CNRS, UPS, France

\vspace{.1cm}
$^{3}$Universit\'e de Lyon, INSA-Lyon, Universit\'e Claude Bernard Lyon 1, UJM-Saint Etienne, CNRS, Inserm, CREATIS UMR 5220, U1206, Villeurbanne, France.

$^{\ast}$Corresponding author: Sayantan~Dutta (Email: sayantan.dutta@irit.fr; sayantan.dutta110@gmail.com).

\vspace*{5mm}
\end{center}

\end{onecolumn}

\section{Introduction}
\label{sec:intro_supp}

\IEEEPARstart{I}{n} this work, we introduce a novel deep-learning (DL) network unfolding the baseline Denoising by Quantum Interactive Patches (De-QuIP) \cite{dutta2021image, dutta2022novel} algorithm, denoted as DIVA (Deep denoising by quantum InteractiVe pAtches) for image denoising problem. We further extend the network architecture to conduct a general image restoration task and the respective network denoted as DIVA advanced (DIVA-A).
The integration of the key attributes of DL and quantum theory significantly enhances the functionality of our proposed networks due to its intrinsic versatility and enables our models to exhibit state-of-the-art performances for several restoration tasks such as denoising, deblurring, super-resolution, inpainting, etc.

In the original manuscript, we extensively study the network architecture and present comprehensive comparisons with benchmark approaches. The detailed quantitative and qualitative analyses are reported in the original manuscript. In this supplementary material, we depict more restored images for the image deblurring, super-resolution, and inpainting problems to give better insights into the visual qualities of the images recovered by our proposed networks.

\section{Experimental Results}
\label{sec:expe_results_supp}

In this section, we analyze the qualitative performance of our proposed networks in various image restoration tasks, such as image deblurring, super-resolution, and inpainting.

\subsection{Quantitative Metrics}

For the purpose of quantitative evalution, the peak-signal-to-noise-ratio (PSNR) and the structural similarity (SSIM)
%\cite{wang2004image}
computed between the true and the restored images were used.

\subsection{Qualitative Image Restoration Results}
\label{sec:results_supp}

% add figures  --------------------------------------
\begin{figure*}[b!]
\begin{centering}
\includegraphics[width=1\textwidth]{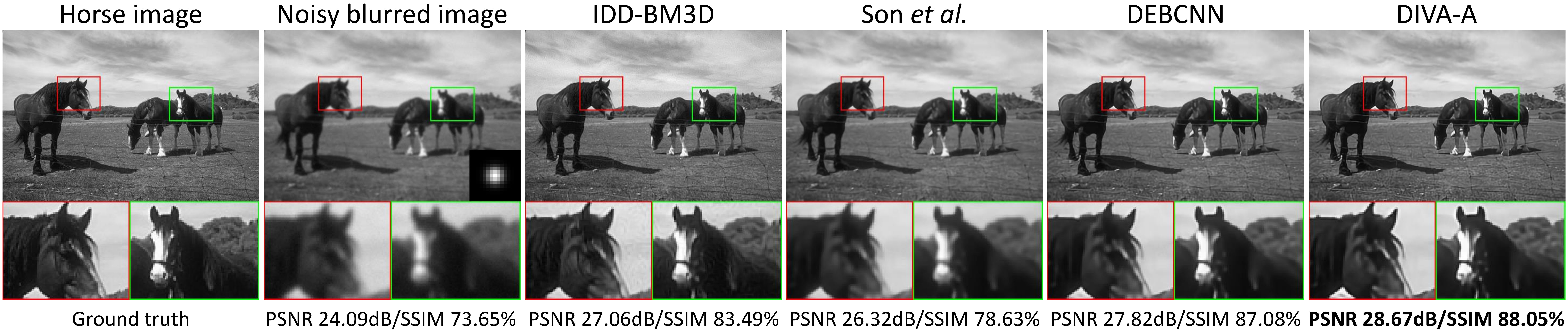}
\end{centering}

\caption{Image deblurring results for \textit{Horse} image degraded by a $25 \times 25$ Gaussian blur kernel of standard deviation 1.6 with random Gaussian noise of standard deviation $2$.}
\label{fig:image_deblurG_supp}
\end{figure*}

% add figures  --------------------------------------
\begin{figure*}[t!]
\begin{centering}
\includegraphics[width=1\textwidth]{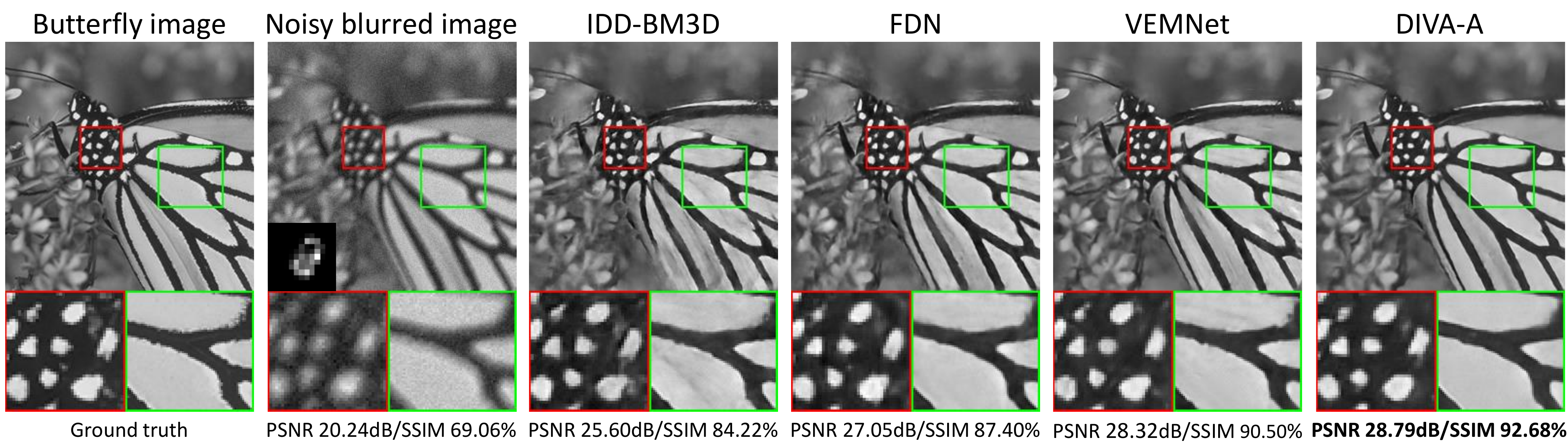}
\end{centering}

\caption{Deblurring results for motion blur kernel. The restored \textit{Butterfly} images with $13 \times 13$ motion blur kernel and Gaussian noise of standard deviation $7.65$.}
\label{fig:image_deblurM1_supp}
\end{figure*}

\begin{figure*}[t!]
\begin{centering}
\includegraphics[width=1\textwidth]{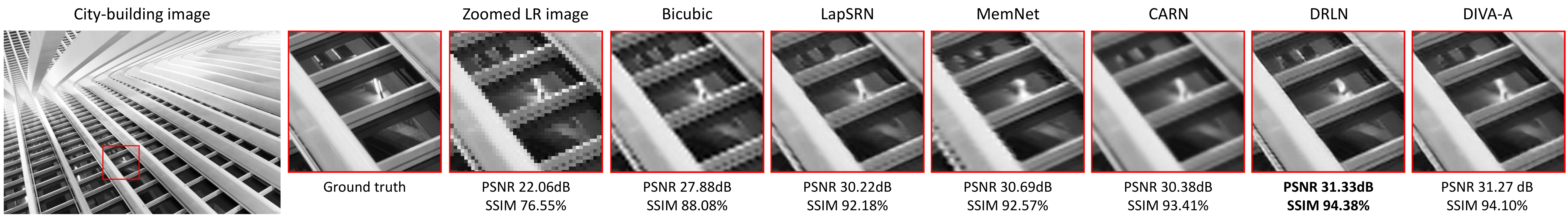}
\end{centering}

\caption{A zoomed regions of the restored HR \textit{City-building} images, extracted from SR results for a bicubic downsampling with scaling factor 3.}
\label{fig:image_super_resoBx3_supp}
\end{figure*}

\subsubsection{Image Deblurring}
\label{sec:result_deblur_supp}

In Fig.~\ref{fig:image_deblurG_supp}, through qualitative evaluation, one can notice that our method not only generates better image contrast but also retrieves sharp edges with more details than other models.  For example, in the \textit{Horse} image, IDD-BM3D \cite{Danielyan2012bm3d} produces better contrast, but random patterns are visible in the deblurred outputs, whereas Son \textit{et al.} \cite{Son2017fast} fails to preserve sharp edges and the overall images appear blurred compared to others. Our DL model restores the head with much sharp and precise edges than the DEBCNN \cite{Wang2018training}, where the edges look hazy. Thus, though DEBCNN \cite{Wang2018training} and our DIVA-A are the two best models in this setting, our model uniformly dominates the sophisticated DEBCNN \cite{Wang2018training} method in the Gaussian deblurring problems.

In Fig.~\ref{fig:image_deblurM1_supp}, the \textit{Butterfly} image is degraded by a moderate size $13 \times 13$ motion blur kernel with random Gaussian noise of standard deviation $7.65$. Our proposed model for retrieving original image quality is significantly better than other competitors. For example, IDD-BM3D \cite{Danielyan2012bm3d}, FDN \cite{Kruse2017learning}, and VEMNet \cite{Nan2020variational} fail to properly restore the pattern on the butterfly's wings and body, and show many distortions in the restored images, as visible in the zoomed boxes. Compared to its rivals, our model has the capability to restore these subtle attributes like patterns on the wings and body, and preserves the sharp edges with finer precision, as shown in the zoomed boxes.

Thus, under both Gaussian and motion blur kernels the overall visual quality of the recovered deblurred images by our proposed model is the best among all the tested methods.

\subsubsection{Single Image Super-Resolution (SR)}
\label{sec:result_sr_supp}

The visual inspections of Figs.~\ref{fig:image_super_resoBx3_supp},~\ref{fig:image_super_resoBx4_supp} and \ref{fig:image_super_resoGx3_supp} illustrate the potential of our method for SR. Figs.~\ref{fig:image_super_resoBx3_supp} and \ref{fig:image_super_resoBx4_supp} display the restored high-resolution (HR) images from the low-resolution (LR) bicubic down-sampled \textit{City-building} and \textit{Fish} images with scale factors of 3 and 4, respectively. The visual effects of HR images recovered by our method are better than others and higher in accuracy. For example, in our retrieved HR \textit{City-building} image the sharp edges of the windows, in \textit{Fish} image the patterns on the fish and the shapes of the seagrass have better specifications than the other methods, such as LapSRN\cite{Lai2017deep}, MemNet\cite{Tai2017MemNet}, CARN\cite{Ahn2018fast}.
Although, the benchmark DRLN \cite{Anwar2022densely} provides better quantitative data in some aspects, the proposed DIVA-A not only gives comparable results but also outperforms DRLN \cite{Anwar2022densely} in terms of quantitative and visual assessments in some cases.

\begin{figure*}[t!]
\begin{centering}
\includegraphics[width=1\textwidth]{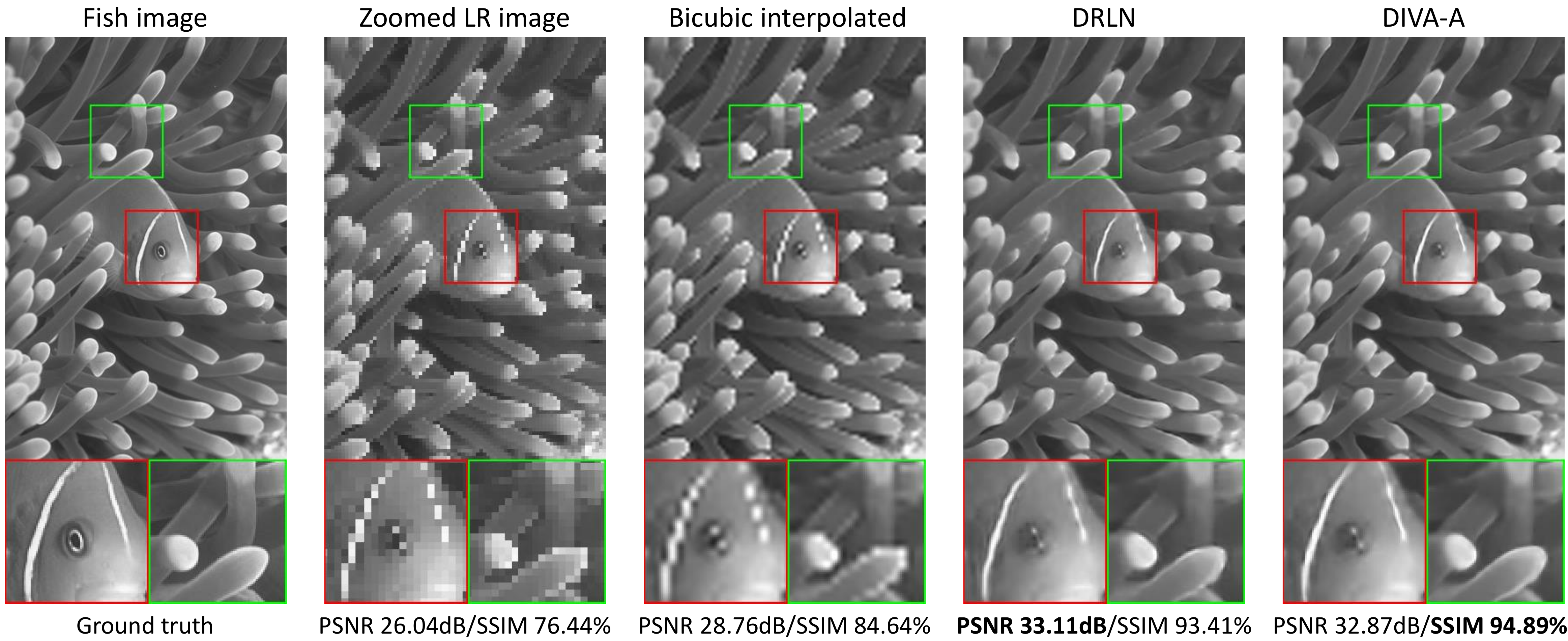}
\end{centering}

\caption{The restored HR \textit{Fish} images from LR images generated by bicubic downsampling with scaling factor 4.}
\label{fig:image_super_resoBx4_supp}

\end{figure*}

% add figures  --------------------------------------
\begin{figure*}[t!]
\begin{centering}
\includegraphics[width=1\textwidth]{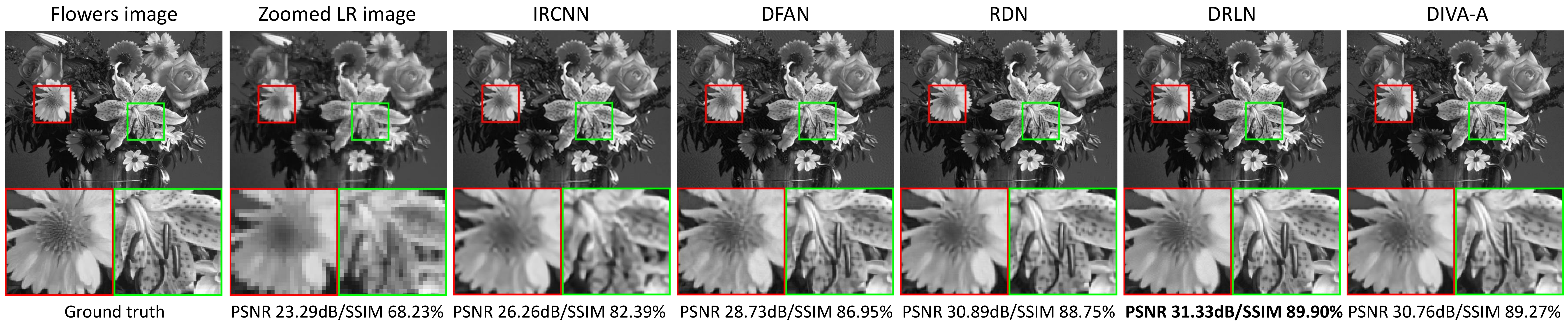}
\end{centering}

\caption{Restored HR \textit{Flowers} images from LR images generated by Gaussian downsampling under a $7 \times 7$ Gaussian blur kernel of standard deviation 1.6 with scaling factor 3.}
\label{fig:image_super_resoGx3_supp}

\end{figure*}

Fig.~\ref{fig:image_super_resoGx3_supp} shows the reconstructed HR images from the LR \textit{Flowers} image obtained by Gaussian downsampling with scale factor of 3. In the degraded image, the small-scale details are nearly unrecognizable. Observation reveals that our method efficiently recovers the edges and patterns of the original image from LR data compared to the state-of-the-art IRCNN \cite{Zhang2017learning}, DFAN \cite{Li2022DFAN} and RDN \cite{Zhang2021residual} methods. Moreover, our method strongly competes with the benchmark DRLN \cite{Anwar2022densely} and beats it in some respects, especially in terms of overall visual quality and preservation of the image structure.

Hence qualitatively, our method is always among the best two approaches in this context. These results can be explained by the ability of DIVA-A to exploit the local structures/attributes via the interaction layer. This layer enables our DL network to efficiently conduct super-resolution tasks while correclty restoring patterns, sharp edges and other small-scale details.

%Thus, harnessing this non-local blueprint, our DL network recovers the HR images with better visual salient attributes like small scale patterns, sharp edges and textures for super-resolution problems.

\subsubsection{Image Inpainting}
\label{sec:result_inpaint_supp}

The visual analysis of Fig.~\ref{fig:image_inpainting_supp} confirms the excellence of our DL model in inpainting tasks. In the \textit{Boat} image, despite 80\% of data missing our model recovers minute details like the ropes and structures on the deck. On the contrary, the image restored by IRCNN \cite{Zhang2017learning} is more blurry and loses/distorts many details, such as the borders, sharp edges, and ropes in the restored output. Hence, our model can gather local information from the image neighborhood quite promisingly through the quantum interaction layer and delivers a high-quality restored image even with limited pixels available.

% add figures  --------------------------------------
\begin{figure*}[t!]
\begin{centering}
\includegraphics[width=.7\textwidth]{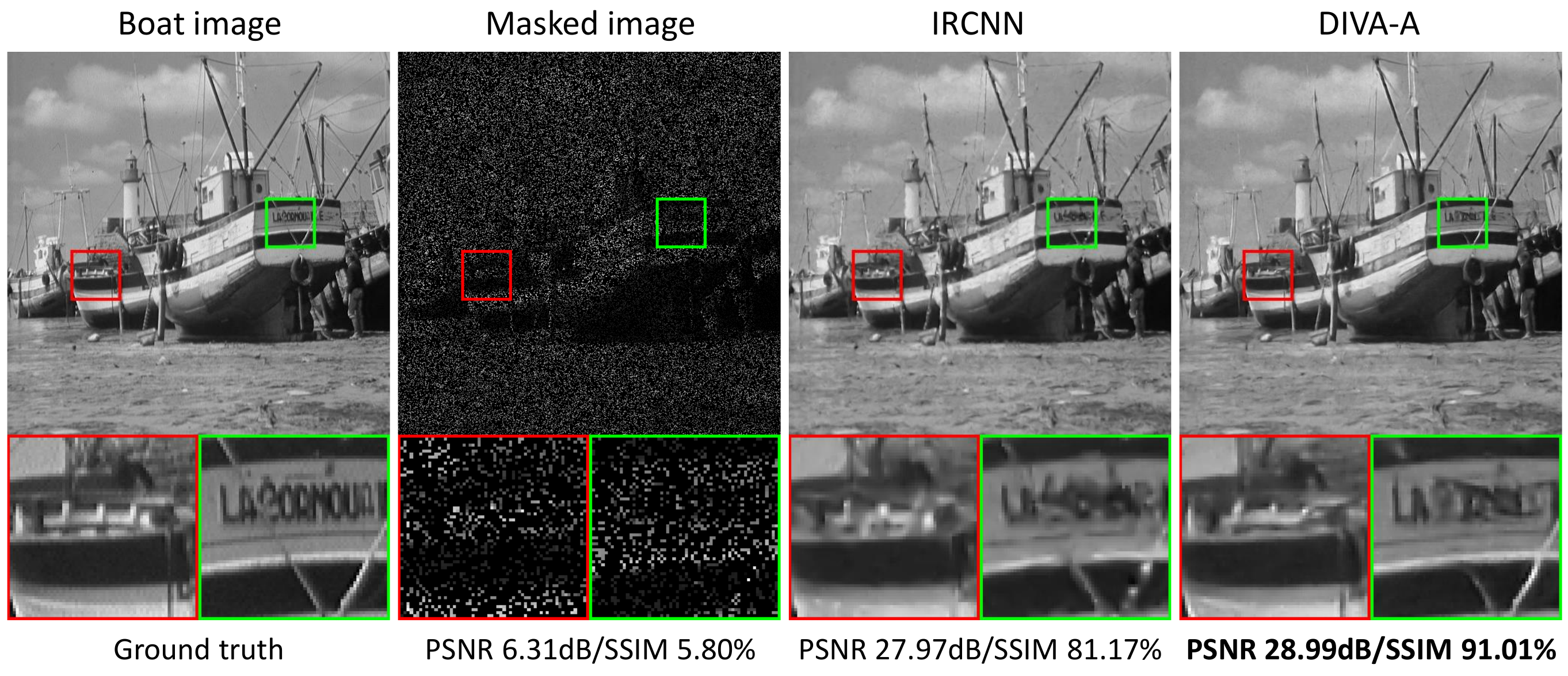}
\end{centering}

\caption{Restored \textit{Boat} images, when 80\% pixels' are missing.}
\label{fig:image_inpainting_supp}

\end{figure*}

\end{document}